\documentclass[fleqn,usenatbib]{mnras}

\usepackage{newtxtext,newtxmath}

\usepackage[T1]{fontenc}
\usepackage{ae,aecompl}

\usepackage{graphicx}	
\usepackage{amsmath}	

\newcommand{\dens}{~g$\cdot$cm$^{-3}$}

\newcommand{\kernelsha}{W(\vert {\bf s}_b-{\bf s}_a\vert,h_a)}
\newcommand{\kernel}{W_{ab}(h_a)}

\title[Axisymmetric  SPMHD]{Axisymmetric  smoothed particle magneto-hydrodynamics}

\author[D. Garc\'\i a-Senz, R. Wissing, R. M. Cabez\'on, E. Vurgun, M. Linares]{
D. Garc\'\i a-Senz $^{1,2}$,\thanks{E-mail: domingo.garcia@upc.edu}
R. Wissing $^3$,
R. M. Cabez\'on$^{4}$,
E. Vurgun$^{1}$, 
M. Linares$^{5,1}$
\\
$^{1}$Departament de F\'\i sica. Universitat Polit\`ecnica de Catalunya. Avinguda Eduard Maristany 16. E-08019 Barcelona (Spain).\\
$^{2}$Institut d'Estudis Espacials de Catalunya. Gran Capit\`a 2-4. E-08034 Barcelona (Spain)\\
$^{3}$Institute of Theoretical Astrophysics. University of Oslo, Postboks 1029, 0315 Oslo (Norway)\\
$^{4}$Center of Scientific Computing -sciCORE, Universit\"at Basel. Klingelbergstrasse 61, 4056 Basel (Switzerland)\\
$^{5}$Department of Physics, Norwegian University of Science and Technology, 7491 Trondheim, (Norway).
}

\date{Accepted XXX. Received YYY; in original form ZZZ}

\pubyear{2015}

\begin{document}
\label{firstpage}
\pagerange{\pageref{firstpage}--\pageref{lastpage}}
\maketitle

\begin{abstract}
Many astrophysical and terrestrial scenarios involving magnetic fields can be approached in axial geometry. Although the smoothed particle hydrodynamics (SPH) technique has been successfully extended to  magneto-hydrodynamics (MHD), a well-verified, axisymmetric MHD scheme based on such technique does not exist yet. In this work we fill that gap in the scientific literature and propose and check a novel axisymmetric MHD hydrodynamic code, that can be applied to physical problems which display the adequate geometry. We show that the hydrodynamic code built following these axisymmetric hypothesis is able to produce similar results than standard 3D-SPMHD codes with equivalent resolution but with much lesser computational load.
\end{abstract}




\section{Introduction}
\label{sec:introduction}
In spite of the large success achieved by Cartesian SPH codes there is a scarcity of SPH calculations taking advantage of the axisymmetric approach in computational fluid dynamics (CFD). To cite a few of them: \cite{herant92}, \cite{petscheck93}, \cite{brook03}, \cite{garciasenz2009}, \cite{shrey19}, \cite{sun21}. Much more dramatic is, however, the case of axisymmetric MHD simulations with SPH (SPMHD) because, as far as we know, there is a manifest void of published material on that topic.

Nevertheless, implementing a consistent, well-verified, axisymmetric SPMHD code may broaden the range of applications of such technique. In astrophysics, the magnetic field around stellar objects can often be described with dipole or toroidal geometries, both consistent with axial geometry. Good examples are the study of magnetized accretion disks around neutron stars and the gravitational collapse of an initially spherical  cloud of a magnetized gas, this last closely related to the formation of proto-planetary disks. Another potential scenario is the core collapse supernova, where magnetic fields and rotation play an important role in the development of the explosion \citep{matsumoto20}. Resolution issues add an extra degree of difficulty when these studies are conducted in three dimensions. In some cases, the axisymmetric approach is the only plausible option to study these scenarios (see, for example, \cite{zanni2009} concerning simulations of accretion onto a dipolar magnetosphere with an Eulerian axisymmetric hydrodynamic code). Furthermore, MHD experiments in terrestrial laboratories can   benefit from the joint virtues of the well-established SPMHD technique \citep{pri08,rosswog2009, price18, wissing20} plus the inherent better resolution of the axisymmetric approach. A paradigmatic example are the Z-pinch devices which aim to focus magnetically driven strong implosions towards the symmetry axis \citep{haines2000}. Additionally, researchers can take advantage of hydrodynamic codes with axial geometry to carry out convergence studies of resolution of their own three-dimensional hydrodynamic codes, or perform computationally affordable parameter explorations.  

In this work we develop and test a novel axisymmetric magneto-hydrodynamic scheme, called Axis-SPHYNX, consistent with the SPH formulation. Our work extends the axisymmetric code developed by \cite{garciasenz2009} to the MHD realm by adding the magnetic-stress tensor to the axisymmetric SPH equations. Furthermore, the induction and dissipative equations are consistently written in such geometry. We focus on the basic mathematical formulation of ideal MHD, so that explicit current terms do not appear in the governing equations. The involved physics is kept as simple as possible: ideal equation of state (EOS), heat transport not included, and no chemical or nuclear reactions. 
We show that, given an axial symmetry, our MHD code is able to produce results similar to those obtained in 3D with SPMHD codes, but with much lesser computational effort. The numerical scheme has been verified with a number of standard tests in ideal MHD, encompassing explosions/implosions, hydrodynamical instabilities, and more complex problems involving self-gravity. 

This paper is organized as follows: section 2 introduces the reader to the axisymmetric formulation of the SPH equations. Such formulation is used to develop a suitable numerical scheme of ideal MHD in Sect.~3. Section~4 is devoted to describe and analyze the results of five numerical tests encompassing a variety of physical scenarios. Finally, a discussion on the results, the conclusions of our work, and future prospects are presented in Sect.~5.   

\section{Axisymmetric formulation of the SPH equations}
\label{sec:idalmhd}

\subsection{Gradient calculation with ISPH}
\label{sec:isph}

Gradients and derivatives are calculated with the Integral Approach (IA) proposed by \cite{garciasenz2012} and adapted to the particularities of axisymmetric geometry. The IA leads to an Integral SPH scheme (ISPH) which was shown to improve the accuracy in estimating  gradients  \citep{cabezon2012,ros15,cabezon17}. Such method is especially suited to handle axisymmetric hydrodynamics, where a good estimation of gradients in points close to the Z-axis is critical. Additionally, the ISPH formalism naturally incorporates corrective terms which are helpful in removing the magnetic tensile-instability. In the IA formalism, the gradient of any scalar function $f$, associated to particle $a$ in the axisymmetric plane, and defined by coordinates ${\mathbf{s}}(r,z)$, with $r=\sqrt{x^2+y^2}$~is,

\begin{equation}
\left[
\begin{array}{c}
\partial f/\partial x^1\\
\partial f/\partial x^2\\

\end{array}
\right]_a
=
\left[
\begin{array}{ccc}
\tau^{11} & \tau^{12}  \\
   \tau^{21}&\tau^{22}
   
\end{array}
\right]_a^{-1}
\left[
\begin{array}{c}
I^1\\
I^2\\
\end{array}
\right]_a\,.
\label{matrix}
\end{equation}

From now on we use the notation $x^1\equiv r; x^2\equiv z; x^3\equiv \varphi$ (with $\varphi$~being the azimuth angle) {\sl indistinctly}\footnote{Note that our index notation slightly differs from that in \cite{garciasenz2012}. Coordinate indexes $\{i,j,k\}$ (as well as $\{r, z,\varphi\}$) are notated superscript to make them compatible to the standard notation of the magnetic-stress tensor. Also note the change in the order at which cylindrical coordinates appear:  $\{r,\varphi,z\}$ in the standard notation, and $\{r, z,\varphi\}$ in this work,  which emphasizes that the axisymmetric plane is mainly defined by the pair $\{r,z\}$.}. Coefficients $\tau^{ij}$ $(i,j=1,2)$, and $I^i$ in Eq.~(\ref{matrix}) are,

\begin{equation}
\tau^{ij}_{a}=\sum_b^{n_b} \frac{m_b}{\eta_b}(x^i_{b}-x^i_{a})(x^j_{b}-x^j_{a})W_{ab}(\vert {\bf s}_b-{\bf s}_a\vert,h_a)\,,
\label{tauijsph}
\end{equation}

\begin{equation}
\begin{split}
    I({\mathbf r}_a)=&\sum_b^{n_b}\frac{m_b}{\eta_b} f({\mathbf r}_b)({\mathbf s}_b-{\mathbf s}_a)~W_{ab}(\vert {\bf s}_b-{\bf s}_a\vert,h_a)\\
    &- f({\mathbf r}_a)\sum_b^{n_b} \frac{m_b}{\eta_b} ({\mathbf s}_b-{\mathbf s}_a) W_{ab}(\vert {\bf s}_b-{\bf s}_a\vert,h_a)\,,
    \end{split}
    \label{iadfull}
\end{equation}

\noindent where $n_b$ is the number of neighbors of the particle, $W_{ab}$ is the kernel function, $h_a, m_b$~are the smoothing length and the  mass of the particle respectively, and $ \eta_b$ is the surface density. The anti-symmetric properties of the gradient of the kernel ensure that the second term in the RHS of Eq.~(\ref{iadfull}) is close to zero. Thus, it is neglected. That assumption  gives rise to the conventional ISPH scheme  \citep{garciasenz2012,ros15}. An exception to that procedure, which is connected with the magnetic tensile-instability problem, is discussed in Sect.~\ref{sec:tensile}. 

From now on, $\kernel\equiv\kernelsha$, with $\vert {\bf s}_b-{\bf s}_a\vert=\sqrt{(r_b-r_a)^2+(z_b-z_a)^2}$, for the sake of clarity. 

\subsection{The Euler hydrodynamic equations in axisymmetric geometry}
\label{sec:axisformulation}

Because the axisymmetric formulation of SPH is probably not too familiar to many readers, we first describe the Euler hydrodynamic equations and discuss the MHD formalism later. The basic Euler ISPH equations in axisymmetric geometry can be directly written from the well known 3D-Cartesian SPH schemes, but changing  the interpolating kernel to $W_{2D}({\bf s})$~ and with the following relationship between the volumetric,  $\rho$, and surface, $\eta$, densities,

\begin{equation}
\rho=\frac{\eta}{2\pi r}\,,
\label{density1}
\end{equation}

\noindent
which evidences that particles are not point-like entities but rings. As a result, the mass of the particles is, in general,  not constant in axisymmetric schemes. The basic axisymmetric Euler equations used in this work  \citep{brookshaw1985,garciasenz2009,relano2012} are shown in Appendix~\ref{sec:appendix A}. These equations are adapted to the IA formalism given by   Eqs.~(\ref{matrix},~\ref{tauijsph}). The derivatives of the kernel are then calculated with \citep{cabezon2012},

\begin{equation}
\frac{\partial {W}_{ab}(h_a)}{\partial x^i_{a}}\Longleftrightarrow {\mathcal A}^i_{ab}(h_a)\,;~ i=1,2\,,
\label{gradk}
\end{equation} 

\noindent with, 

\begin{equation}
    \mathcal A^i_{ab}(h_{a,b})=\sum^2_{j=1} c^{ij}_{a}(h_a)(x^j_{b}-x^j_{a})W_{ab}(h_{a,b})
    \,,
\end{equation} 

\noindent being $c^{ij}_{a}$ the coefficients of the inverse matrix in the IA given by Eq.~(\ref{matrix}). 

\begin{figure}
\includegraphics[width=0.47\textwidth]{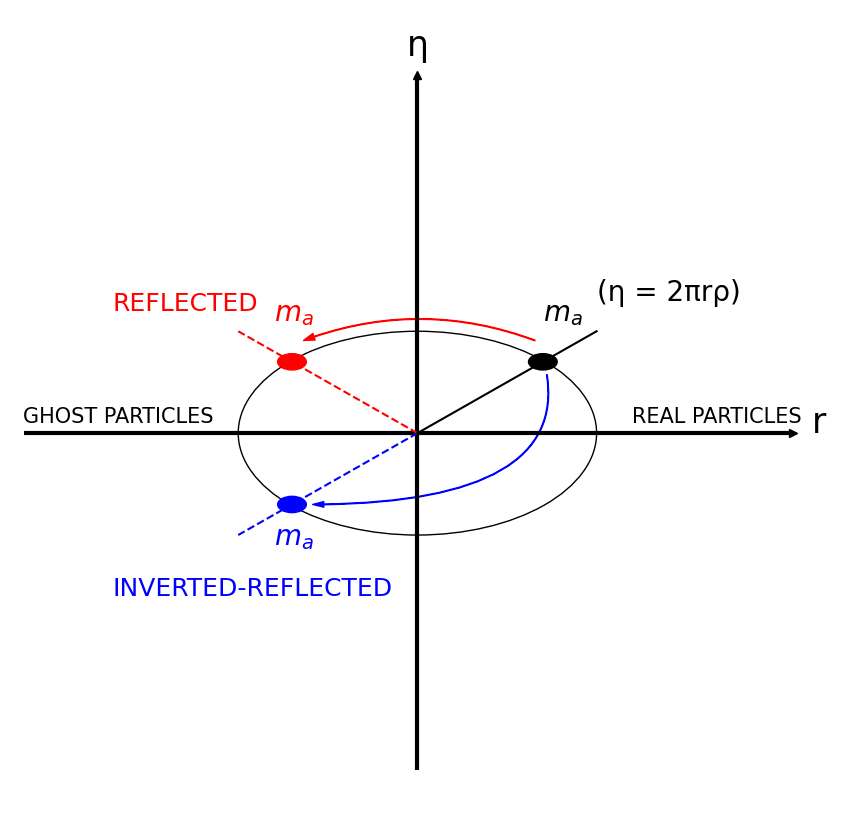}
\caption{The use of inverted-reflected ghost particles along with the IA method trivially avoids the singularity problems when calculating the density $\eta_a$ near the symmetry axis. We show here surface density in the Y-axis, hence inverted-reflected ghost particles have negative $\eta$.}
\label{fig:invrefl}
\end{figure}

We stress that although the main Axis-SPH equations are henceforth written within the ISPH formalism, translating them to the standard SPH scheme with expression~\ref{gradk} is straightforward  ( a calculation with traditional derivatives is shown in  Sect.~\ref{subsec:sedov}). According to Appendix~\ref{sec:appendix A}, the Axis-SPH equations are as follows:

\vskip 0.1 cm
\begin{itemize}
\item {\sl Mass equation},
\begin{equation}
\eta_a=\sum_{b=1}^{n_b} \varepsilon_b~ m_b W_{ab}(h_a)\,.
\label{surfacedensity}
\end{equation}
\vskip 0.2 cm

\item {\sl Momentum equations},
\begin{equation}
\begin{split}
a^r_a =~ &2\pi\frac{P_a}{\eta_a}-\\
&2\pi \sum_{b=1}^{n_b}  m_b \left(\frac{\varepsilon_{b,1} P_a \vert r_a\vert}{\eta_a^{2-\sigma}\eta_b^{\sigma}}{\mathcal A}^r_{ab}(h_a)+
 \frac{\varepsilon_{b,2} P_b \vert r_b\vert}{\eta_b^{2-\sigma}\eta_a^{\sigma}} {\mathcal A}^r_{ab}(h_b)\right)\,,
\end{split}
\label{accel_r}
\end{equation}

\begin{equation}
a^z_a = -2\pi
\sum_{b=1}^{n_b} m_b\left(\frac{\varepsilon_{b,1} P_a \vert r_a\vert}{\eta_a^{2-\sigma}\eta_b^{\sigma}}{\mathcal A}^z_{ab}(h_a)+\frac{\varepsilon_{b,2} P_b \vert r_b\vert}{\eta_b^{2-\sigma}\eta_a^{\sigma}}{\mathcal A}^z_{ab}(h_b)\right)\,.
\label{accel_z}
\end{equation}

\item {\sl Energy equation},

\begin{equation}
\begin{split}
\frac{du_a}{dt} =-&2\pi\frac{P_a}{\eta_a} v_{r_a}+\\
&2\pi\frac{P_a \vert r_a\vert}{\eta_a^{2-\sigma}}\sum_{b=1}^{n_b}\frac{\varepsilon_{b,1}}{\eta_b^{\sigma}} m_b \left(v^i_{ab}~{{\bf\mathcal A}^i_{ab}}(h_a)\right)\,,
\label{energy1}
\end{split}
\end{equation}

\noindent where $P_{a,b}, u_a$~are the pressure and specific internal energy, and $v^i_{ab}=v_a^i-v_b^i$. The binary parameter $\sigma[0,1]$ allows to choose between the two most widely used SPH schemes (see Appendix~\ref{sec:appendix A}),  

\begin{equation}
\mathrm \sigma =
\left\{\begin{array}{rclcc}
0 & \qquad\mathrm{Euler-Lagrange~schemes} \\
1 & \qquad\mathrm{geometric-density~averaged~schemes}, 
\end{array}
\right.\,.
\label{scheme}
\end{equation}

\noindent and the meanings of $\varepsilon_b$, $\varepsilon_{b,1}$ and $\varepsilon_{b,2}$ are commented below. Axisymmetric SPH schemes arising from the Euler-Lagrange equations ($\sigma=0$~in Eq.~\ref{scheme}) were discussed in
\cite{brookshaw1985, garciasenz2009} and \cite{shrey19}. On another note, Cartesian SPH schemes built with $\sigma=1$ are more effective in suppressing the tensile instability than schemes with $\sigma=0$ \citep{rea10, Wadsley17}. It is also feasible to make use of an adaptive sigma, so that the scheme is Lagrangian compatible in a large  fraction of the system \citep{gsenz22}. In this work we focus on the crossed-density scheme $\sigma=1$, because not only removes the tensile instability, but allows a direct comparison with the results by \cite{wissing20} in the verification tests in Sect.~\ref{sec:Tests}.  

The parameter $\varepsilon_b$ in Eq.~(\ref{surfacedensity}) (see also Table~\ref{tab:table1}) is, 

\begin{equation}
\mathrm \varepsilon_{b}=
\left\{\begin{array}{rclcc}
+1 & \qquad\mathrm{Real~particles}\,, \\
-1 & \qquad\mathrm{Axis-ghost~ particles}, 
\end{array}
\right.
\label{signature}
\end{equation}

\noindent which assigns a signature to the neighbor particle. According to the discussion below, the introduction of the sign $\varepsilon_b$~in the scheme ensures that $\eta_a$~is correctly calculated with Eq.~(\ref{surfacedensity}) in the proximity of the singular axis. 

The set of SPH equations above differs from those arising from a 2D-Cartesian description in several ways. Firstly, there are the first terms on the RHS of Eqs.~(\ref{accel_r}) and (\ref{energy1}), which are called {\sl hoop-stress} terms. These are specific of the axisymmetric formulation. Another particularity are the multiplicative $\vert r_a\vert, \vert r_b\vert$ elements appearing inside the summations. As shown in Appendix \ref{sec:appendix A}, these come after inverting the volumetric density in the Euler equations.  Finally,  there is a difference in the treatment of the particles moving around the singular axis $Z$. Close to the Z-axis, the cylindrical symmetry enforces $\rho_{r\to 0}=\rho_0$ and therefore $\eta=2\pi\vert r\vert\rho_0\rightarrow 0$ a feature which is not guaranteed when simple reflective ghost particles are used. Such unwanted behavior can be cured by multiplying $\eta$ and ${\bf \nabla}\eta$ by a corrective factor, so that the limit above is enforced \citep{garciasenz2009}. Another solution, sketched in Fig.~\ref{fig:invrefl}, is to compute the contribution to surface density of ghost particles as having negative density (inverted-reflected particles). According to Fig.~\ref{fig:invrefl}, this recipe restores the linearity of $\eta(r)$, leading to exact interpolations close to the symmetry axis when Eqs.~(\ref{surfacedensity}) and (\ref{signature}) are used to compute the surface density. A basic feature of ISPH is that the gradient of the surface density of a  particle is determined comparing the values of $\eta$~within the cluster of neighboring particles. Thus, a good depiction of $\eta(r\rightarrow 0)$ guarantees that ${\bf\nabla}\eta(r\rightarrow 0)$ is well evaluated when the integral approach, Eq.~(\ref{matrix}) is used to compute the gradient. 
 
On another note, including sign-axis corrections in the momentum and energy equations is also necessary, and it improved the results in the studied test cases. The occurrence of $\varepsilon_b$ in these equations is due to the fact that inverted-reflected ghost particles have $\{m_b<0;~ r_b < 0;~\eta_b<0\}$. Because equations (\ref{accel_r}), (\ref{accel_z}), and (\ref{energy1}) work with positive masses ($m_b$), radial distances ($\vert r_a\vert, \vert r_b\vert$), and surface densities ($\eta_a,\eta_b$), we need to include the signature $\varepsilon_b$ to account for the axis-ghost particles via the parameters $\varepsilon_{b,1}$ and $\varepsilon_{b,2}$, as shown in these equations. The value of  $\varepsilon_{b,1}; \varepsilon_{b,2}$ in the Axis-SPH equations, as a function of the chosen SPH scheme, $\sigma$, is shown in Table~\ref{tab:table1}.   
 
\begin{table}
	\centering
	\caption{Sign of axis-ghost particles in Eqs.~(\ref{surfacedensity}, \ref{accel_r}, \ref{accel_z}, \ref{energy1}) as a function of the chosen SPH scheme, determined by $\sigma$. Real particles have $\varepsilon_b=\varepsilon_{b,1}=\varepsilon_{b,2}=1$.}
	\begin{tabular}{lcccccr} 
		\hline
		Scheme &$\varepsilon_b$& $\varepsilon_{b,1}$&$\varepsilon_{b,2}$ \\
		\hline
		$\sigma=0$&$-1$&$-1$&$+1$\\
		$\sigma=1$&$-1$&$+1$&$-1$ \\
		\hline
	\end{tabular}
	\label{tab:table1}
\end{table}

\end{itemize}

\section{Formulation of ideal MHD in axial geometry}
\label{sec:isphmhd}

Adapting the axisymmetric ISPH equations to MHD is not too complicated. The mass-equation, Eq.~(\ref{surfacedensity}), does not change. In the momentum equations, Eqs.~(\ref{accel_r}) and (\ref{accel_z}), the pressure terms are replaced by the magnetic stress tensor \citep{pri12}, 

\begin{equation}
    S_a^{ij}=-\left(P_a+\frac{1}{2\mu_0}B_a^2\right)\delta^{ij}+\frac{1}{\mu_0}\left(B_a^iB_a^j\right)\,,
    \label{mhdstress}
\end{equation}

\noindent
where letters subscripts, $\{a,b\}$, refer to particles, and $\{i=1,3;~j=1,3\}$ are tensor components. Even though the scheme is basically two-dimensional, with coordinates ${\bf s}(r,z)$, a third coordinate, associated to the azimuth angle $\varphi$, could be eventually necessary to describe the toroidal component of the magnetic field and velocity. These momentum equations must also include the magnetic contribution to the hoop-stress term. The derivation of the axisymmetric SPMHD equations, using the least action principle \citep{pri12}, is shown in Appendix \ref{sec:appendix B}. The axisymmetric equation of energy, Eq.~(\ref{energy}), remains unchanged.

We write the axisymmetric SPMHD scheme only in its density averaged, ``crossed'',  form (i.e. $\sigma=1$), because these are the equations used in this work.    
\newline
\newline
\begin{itemize}
    \item {\sl Mass conservation}
    
\begin{equation}
\eta_a=\sum_{b=1}^{n_b} \varepsilon_b~m_b W_{ab}(h_a)
\label{densitymhd}
\end{equation}
\newline
   \item {\sl Momentum equations}
   
   \begin{equation}
\begin{split}
a^r_a=~ & 2\pi\frac{\left(P_a+\frac{B_a^2}{2\mu_0}-\frac{(B_a^\varphi)^2}{\mu_0}\right)}{\eta_a}+\\
&2\pi \sum_{b=1}^{n_b}  m_b \left(\frac{S_a^{ri} \vert r_a\vert}{\eta_a\eta_b}{\mathcal A}_{ab}^{i}(h_a)+
 \varepsilon_b\frac{S_b^{ri} \vert r_b\vert}{\eta_a\eta_b} {\mathcal A}_{ab}^{i}(h_b)\right)\,, 
\end{split}
\label{mhdaccel_r}
\end{equation}

\begin{equation}
a^z_a=  2\pi\sum_{b=1}^{n_b} m_b\left(\frac{S_a^{zi} \vert r_a\vert}{\eta_a\eta_b}{\mathcal A}_{ab}^{i}(h_a)+\varepsilon_b\frac{S_b^{zi} \vert r_b\vert}{\eta_a\eta_b}{\mathcal A}_{ab}^{i}(h_b)\right)\,,
\label{mhdaccel_z}
\end{equation}

\begin{equation}
\begin{split}
a^\varphi_a&=  2\pi\left(\frac{B_a^r B_a^{\varphi}}{\mu_0\eta_a}\right)+\\
&2\pi \sum_{b=1}^{n_b}  m_b \left(\frac{S_a^{\varphi i} \vert r_a\vert}{\eta_a\eta_b}{\mathcal A}_{ab}^{i}(h_a)+
 \varepsilon_b\frac{S_b^{\varphi i} \vert r_b\vert}{\eta_a\eta_b} {\mathcal A}_{ab}^{i}(h_b)\right)\,,
\end{split}
\label{mhdaccel_phi}
\end{equation}

\noindent where, repeated indexes in $\{i=r,z\}$~are summed. Equation (\ref{mhdaccel_phi}) is only relevant in those applications involving \{$v_\varphi,B_\varphi\neq 0$\}, as it is the case of scenarios combining rotation and toroidal magnetic fields. Its impact in the simulations is discussed in Sect.~\ref{subsec:collapse}.
\newline
\item {\sl Energy equation}

\begin{equation}
\begin{split}
\frac{du_a}{dt}=-2\pi\frac{P_a}{\eta_a} v_{r_a}+
&2\pi\frac{P_a \vert r_a\vert }{\eta_a}\sum_{b=1}^{n_b} \frac{m_b}{\eta_b} \left( v^i_{ab}~{{\bf\mathcal A}^i_{ab}}(h_a)\right)\,.
\label{energy}
\end{split}
\end{equation}

\end{itemize}

\subsection{The induction equation}

First, we write the induction equation in a similar manner as in \cite{pri12},

\begin{equation}
\frac{d{\bf B}}{dt} = -{\bf B}({\bf\nabla}\cdot{\bf v}) + ({\bf B}\cdot {\bf \nabla}){\bf v}\,,
\label{induction}
\end{equation}

\noindent
where the non-ideal term associated with the current density $\bf J$ has been removed from the expression. Secondly, we write ${\bf B}({\bf\nabla}\cdot{\bf v})$ and the material derivative $({\bf B}\cdot {\bf \nabla}){\bf v}$ in cylindrical coordinates, assuming $\frac{\partial}{\partial \varphi}=0$. Finally, we have,

\begin{equation}
 \begin{split}
 \frac{d}{dt} 
 &
\left[
\begin{array}{c}
 B^r \\ 
  B^z \\
 B^\varphi 
\end{array}
\right]
= \\
    &\left[
\begin{array}{ccccc}
-\left(\frac{\partial v^z}{\partial z}+\frac{v^r}{r}\right) & \frac{\partial v^r}{\partial z}& -\frac{v^\varphi}{r} \\
\frac{\partial v^z}{\partial r}& -\left(\frac{\partial v^r}{\partial r}+\frac{v^r}{r}\right) & 0\\
\frac{\partial v^\varphi}{\partial r}&\frac{\partial v^\varphi}{\partial z} &-\left(\frac{\partial v^r}{\partial r}+\frac{\partial v^z}{\partial z}\right) \\
\end{array}
\right]
\left[
\begin{array}{c}
 B^r \\ 
  B^z \\
 B^\varphi
\end{array}
\right]\,.
\end{split}
\label{matrixind}
\end{equation}  

Thus, the induction equation is expressed as a linear equation,

\begin{equation}
\frac{d{B^i_{a}}}{dt} = \sum_{j=1}^3 r^{ij} B^j_{a}
\label{linearinduction}
\end{equation}

\noindent
where $B^i_{a}$~is the $i-$component of the magnetic field acting on  particle $a$, and coefficients $r^{ij}$~ only depend on the velocity field around the particle\footnote{The  induction equation, Eq.~(\ref{linearinduction}), has been integrated explicitly in this work. Nonetheless, it can also  be approached implicitly by inverting the matrix in Eq.~(\ref{matrixind}). An implicit solver could be appropriate in those cases where, for whatever reason, the system of differential equations governing the induction equation become stiff.}.

\subsection{Dissipation}

As in Cartesian SPH, the axisymmetric approach demands some amount of dissipation to handle shock waves. As it is usual in SPH, this is done with the artificial viscosity (AV) concept. There are two main sources of dissipation in MHD: those from the AV and those arising from the induced currents in plasma sheets during collisions. The former is purely hydrodynamical and is that implemented in Axis-SPHYNX and described in \cite{cabezon17} with  the third spatial component removed. For the latter, we use the scheme described in \cite{tricco13,price18,wissing20}. We show here both for completeness. The viscous acceleration is written as,   

\begin{equation}
\begin{aligned}
a^{i,AV}_{a}= -\frac{1}{2m_a}\sum_b &\left\{V_a~m_b~\Pi'_{ab}~f_a~A^i_{ab}(h_a)\right.\\ 
+&~\left.V_b~m_a~\Pi'_{ab}~f_b~A^i_{ab}(h_b)\right\}\,,
\label{accvis_2}
\end{aligned}
\end{equation} 

\noindent
with, 

\begin{equation}
\Pi'_{ab}=
\begin{cases}
-\frac{\alpha}{2}~v_{ab}^{sig}~w_{ab} & \text{for~~$\mathbf{r}_{ab}\cdot\mathbf{v}_{ab} < 0$} \\
 0 & \text{otherwise}\,,
\end{cases}
\label{avis_2}\,
\end{equation}

\noindent
where $V_a= m_a/\eta_a$~is the 2D volume element and $f_a, f_b$ are the Balsara limiters \citep{bal95}:

\begin{equation}
f_a=\frac{\vert\nabla\cdot\mathbf{v}\vert}{\vert\nabla\cdot\mathbf{v}\vert+\vert\nabla\times\mathbf{v}\vert+10^{-4}~c_a/h_a}\,.
\label{balsara}
\end{equation}

The signal velocity includes a quadratic term, which is adequate to handle strong shocks \citep{price18},   

\begin{equation}
    v_{ab}^{sig} = \alpha\bar c_{ab,s}-\beta w_{ab}
\end{equation}

\noindent where $w_{ab}={\bf v}_{ab}\cdot {\bf \hat r}_{ab}$~and $\bar c_{ab,s}$~is the average of the sound speed between particles $a,b$. The parameters $\alpha$ and $\beta$ are kept constant with default values  $\alpha=1$ and $\beta=2$. Future developments of Axis-SPHYNX will incorporate AV switches \citep{cul10, rea10} to better control the dissipation.  

Regarding the magnetic dissipation, some amount is necessary to smooth the transverse component of the magnetic field in shocks. The adopted scheme was that described in \cite{tricco13, wissing20},  

\begin{equation}
    \left(\frac{d\mathbf B}{dt}\right)^{diss}=\xi_B \nabla^2\mathbf B\,,
    \label{Bdiss_1}
\end{equation}

\noindent with $\xi_B = \alpha_B~v_{sig,B}~h$, mimicking a magnetic resistivity coefficient. $v_{sig,B}$ is the characteristic signal velocity and $\alpha_{B}\simeq 1$. The numerical analog of Eq.~(\ref{Bdiss_1}) is rather complicated, hence we show it in Appendix~\ref{sec:appendix C}. It contains a Cartesian part (but with coordinates $r$ and $z$),

\begin{equation}
    \left(\frac{d\mathbf B}{dt}\right)_a^{diss,C}=\sum_{b=1}^{n_b}\frac{m_b}{\eta_b} \frac{\xi_{B,a}+\xi_{B,b}}{\vert s_{ab}\vert}\mathbf B_{ab}\left(\hat s^i_{ab}\tilde A^i_{ab}\right)\,,
    \label{Bdiss_2}
\end{equation}

\noindent where ${\mathbf B}_{ab}={\mathbf B}_a-{\mathbf B}_b$, $\hat s_{ab}$ is the unit vector joining the particles $a,b$ in the axisymmetric plane, and

\begin{equation}
    \widetilde A^i_{ab}= \frac{1}{2}\left[\mathcal A^i_{ab}(h_a)+\mathcal A^i_{ab}(h_b)\right]\,.
    \label{averagedA}
    \end{equation}

In cylindrical geometry, however, there are other contributions (see Appendix~\ref{sec:appendix C}) to be added to the Cartesian part of Eq.~(\ref{Bdiss_1}). The complete expression giving the evolution of each component of the magnetic field is, 

\begin{equation}
   \left(\frac{dB^i}{dt}\right)_a^{diss} = \left(\frac{d B^i}{dt}\right)_a^{diss,C}+\left(\frac{\xi_B}{r}~\frac{\partial B^i}{\partial r}\right)_a - (1-\delta^{iz}) \left(\frac{\xi_B }{r^2}B^i\right)_a\,,
   \label{Bdiss_3}
\end{equation}

\noindent where $\delta^{iz}$ is the Kronecker-delta function. 

The magnetic dissipation contributes to the rate of change of the internal energy. The simplest way to estimate such contribution is to neglect the non-Cartesian part of the dissipation, because it is usually very subdominant. In that case, it is enough to use the expression by \cite{price18} and \cite{wissing20}, but restricted to the axisymmetric plane $\{r,z\}$ \citep{gsenz2022}. A more general procedure to build an energy equation, which takes into account to all terms in Eqs.~(\ref{Bdiss_1}) and (\ref{Bdiss_3}) is to consider, 

\begin{equation}
    \rho\left(\frac{du}{dt}\right)_a^{diss}= \xi_B~  {\mathbf J}\cdot {\mathbf J}\,,
    \label{joule}
\end{equation}

\noindent where ${\mathbf J}=({\mathbf \nabla}\times {\mathbf B})/\mu_0$ is the electric current density vector. Equation~(\ref{joule}) is simply governing the rate of heat (Joule-like) losses per unit mass, and it is a positive definite magnitude. In axial geometry, the components of ${\mathbf J}$ are, 

\begin{equation}
{\mathbf J}=\frac{1}{\mu_0}\left\{-\frac{\partial B^\varphi}{\partial{z}} \hat r+\left(\frac{\partial B^\varphi}{\partial r}+\frac{B^\varphi}{r}\right) \hat z+ \left(\frac{\partial B^r}{\partial z}-\frac{\partial B^z}{\partial r}\right)\hat \varphi\right\}\,,
\end{equation}

\noindent where the derivatives are calculated with the standard SPH procedure. In practice, it is preferable to evaluate the combination ${\mathbf J}_\xi=\xi^{\frac{1}{2}} {\mathbf J}=({\mathbf \nabla}\times {\mathbf \xi^{\frac{1}{2}}B})/\mu_0$, rather than ${\mathbf J}$, in the SPH summations,  because the resistivity is usually defined on pairwise basis, $\xi_B(ab)$. The dissipation rate of magnetic energy is therefore $({\mathbf J}_\xi\cdot {\mathbf J}_\xi)/\rho$.

In the tests below, the adopted value of $\xi_B$ is, 
 
 \begin{equation}
     \xi_B=\frac{1}{2}\alpha_B~v_{sig,B}\vert s_{ab}\vert\,.
 \end{equation}
    
For the signal velocity,  $v_{sig}$, we used the expression by \cite{price18} in most of the tests below, 

\begin{equation}
    v_{sig,B}=\vert \mathbf v_{ab}\times \hat{\mathbf s}_{ab}\vert
    \label{Bdiss_4}\,,
\end{equation}

This showed to produce lesser dissipation than the Alfven velocity, $v_{sig,B}=v_{\mathrm Alfven}=\sqrt{B^2/(\mu_0\rho)}$ far from shocks \citep{price18}. 
    
\subsection{Cleaning the div B}
\label{sec:cleaning}

A big challenge of numerical MHD is to permanently fulfill the Maxwell equation $\mathbf\nabla\cdot \mathbf B = 0$. In most of existing the SPH codes this is achieved with divergence cleaning techniques. Here we use the hyperbolic/parabolic cleaning scheme by \cite{tricco2016} which has proven to be very satisfactory keeping div~$\mathbf B$ at acceptable levels \citep{price18}. Additionally, the method is robust and easy to implement. Adapting such parabolic cleaning scheme to the axisymmetric geometry is straightforward. Basically, a term  $(d\mathbf B/dt)_\psi$~is added to the induction equation, Eq.~(\ref{induction}), so that the magnetic field diffuses and non-zero divergence values are rapidly smeared through the whole system. The $i-$component of that contribution is. 

\begin{equation}
    \left(\frac{d B^i}{dt}\right)_{\psi,a}=-\sum_b \frac{m_b}{\eta_b} \left(\psi_a+\psi_b\right)\mathcal{\tilde   A}^i_{ab}\qquad (i=1,2)\,,
    \label{cleaning_1}
\end{equation}

\noindent where the coefficients $\psi$ evolve following the differential equation \citep{tricco12},

\begin{equation}
    \frac{d}{dt}\left(\frac{\psi}{c_h}\right)=-c_h \mathbf\nabla\cdot\mathbf B - \frac{1}{\tau_h}\frac{\psi}{c_h}-\frac{1}{2}\frac{\psi}{c_h}\mathbf\nabla\cdot\mathbf v\,,
    \label{cleaning_2}
\end{equation}

\noindent and $c_h= f_{\mathrm{clean}}~v_{\mathrm{mhd}}$, with  $v_{\mathrm{mhd}}=\sqrt{c_s^2+v_{Alfven}^2}$ and $\tau_a=h_a/(c_{h,a}
\sigma_c)$ is a relaxation time. Following \cite{wissing20}, the free parameters in the expressions above were set to $f_{\mathrm{clean}}=1$ and $\sigma_c=1$.  

\subsection{Magnetic tensile instability}
\label{sec:tensile}

Calculations where magnetic pressure largely exceeds the kinetic gas pressure are prone to undergo the tensile instability \citep{phillips85}. Such instability concerns the harmful effect of the magnetic-stress tensor elements $B^iB^j/\mu_0$, when they become large enough. The  tensile instability induces strong particle clustering which leads to numerical troubles, especially  when $\vert \mathrm{div}~\mathbf B\vert$~is large.  
One of the firsts solutions to the tensile-instability problem was suggested by \cite{morris96}, who  subtracted the last term in the RHS in Eq.~(\ref{mhdstress}) from the acceleration equation, Eqs~(\ref{mhdaccel_r},\ref{mhdaccel_z}). Commonly used forms of such corrective term to the momentum equation can be found in \cite{borve01}~and \cite{pri12}.

It is worth to note that the ISPH scheme provides naturally a similar  corrective term to that by \cite{morris96}.  According to \cite{garciasenz2012} such term, $f^i_{divB,a}$ is, 

\begin{equation}
 f^i_{divB,a} = - \frac{2}{\mu_0}\sum_b m_b \frac{(B^iB^j)_a}{\rho_a\rho_b}{\nabla_a^j W}_{ab}(h_a)\,, 
 \label{tensile_1}
\end{equation}

The corrective term  is applied wherever the magnetic pressure exceeds the gas pressure ($\frac{B^2}{2\mu_0}> P$). To smooth the transition between the weak and strong field regimes we use the interpolating function by \cite{wissing20}, 

\begin{equation}
{\mathcal H}_{a}=
\left\{\begin{array}{rclcc}
2 & \qquad \beta_a < 1 \\
2(2-\beta_a) & \qquad 1\le\beta_a \le 2\\
0 & \qquad\mathrm {Otherwise}, 
\end{array}
\right.
\label{betainterp}
\end{equation}

\noindent with $\beta_a=\frac{2\mu_0 P_a}{B_a^2}$. Equation~(\ref{tensile_1}), is easily adapted to the  axial-ISPH formalism,

\begin{equation}
 f^i_{divB,a} = -\frac{2\pi \mathcal{H}_a}{\mu_0}\sum_b m_b \frac{(B^iB^j)_a~\vert r_a\vert}{\eta_a\eta_b}{{\mathcal{A}}}^j_{ab}(h_a)\,. 
 \label{tensile_2}
\end{equation}

 The magnitude $f^i_{divB,a}$~in Eq.~(\ref{tensile_2}) is added to Eqs.~(\ref{mhdaccel_r}) and (\ref{mhdaccel_z}) to obtain the acceleration of the SPH-particles.

\subsection{Boundaries}
\label{sec:boundaries}

Arranging boundary conditions in axisymmetric geometry is delicate. On one hand, the $\propto\frac{1}{r}$~dependence of divergence-like expressions which often appear in cylindrical geometry, makes the Z-axis singular. On the other hand, considering ghost particles across the Z-axis is necessary to adequately reproduce the surface density in the axis neighborhood. Adding reflective ghost particles ($r\rightarrow -r$, $z\rightarrow z$, $v_r\rightarrow v_r$, etc) is probably the best option but it has two shortcomings. The first is that the surface density, $\eta$, is not correctly reproduced when $r\rightarrow 0$~(see Sect.~\ref{sec:axisformulation}). The second is that particle penetration through the axis line is not completely avoided. 

Interpolating kernel functions  with cylindric geometry can be used \citep{Omang05} to overcome the first problem above. Another option is to apply a suitable correction function to $\eta(r\rightarrow 0)$, as in \cite{garciasenz2009}. Particle excursions to the negative region, $r<0$, of the plane can be blocked with the addition of ad-hoc repulsive damping forces in the axis neighborhood \citep{Li2020}. 

In this work, we used common reflective ghosts particles, with the exception of the surface density, for which we have introduced the notion of inverted-reflected ghost particles to exactly reproduce $\eta$ when $r\rightarrow 0$ (see Fig.~\ref{fig:invrefl}). The introduction of  inverted-reflected particles makes the profile of $\eta (r\rightarrow 0)$ linear, so that interpolations are exact owing to the second-order accuracy of the SPH technique. Furthermore, the chances of a SPH particle crossing the singularity axis are greatly reduced when considering the arithmetic average of the radial velocity of a particle, $v_a^r$, moving close to the Z-axis,

\begin{equation}
v_a^r\left(\frac{r_a}{h_a} < 2\right) \longrightarrow \left<v_a^{r}\right> = \frac{1}{n_b}\sum_{b} v_b^r\,.
\label{vr_average}
\end{equation}

Replacing $v^r$~by its average, if $r/h<2$, enforces the correct limit of radial velocity, $< v_a^r (r\simeq 0)>\rightarrow 0$, and largely overcomes particle penetration through the Z-axis. A similar recipe can be used to smooth other magnitudes, as for example the r-component of the magnetic field $B_r$.

Periodic boundary conditions are used on the top and bottom sides of the cylinder, while reflective ghost have been used on the lateral surface. Small variations of these default boundary conditions are explicitly stated in some of the tests below.     

\subsection{ Conservation properties} 
\label{sec:conservation}
The formulation of the  SPMHD technique is essentially  conservative. Conservation of momentum and energy is, however, not complete in the strong field regime due to the collateral effect of the  $f_{divB}$ correction \citep{price18}, which is needed to elude the onset of tensile instability. 

The conservation properties of axisymmetric SPH codes are not as good as those shown by Cartesian formulations of SPH. The conservation of linear, angular momentum, and energy in the real semi-plane, ($r\ge 0$), is not perfect. First, there is an exchange of momentum and energy across the Z-axis with the mirror ghost particles. Second, and more important, the hoop-stress term in the momentum equation (first term in the RHS of Eq.~\ref{mhdaccel_r}), does not preserve the linear momentum in the $r-$direction. Nevertheless, when the whole plane [$-r,+r$] is taken into account, the contributions of the hoop-stress force on both sides balance out and linear and angular momentum are in fact conserved. In the tests presented below, the total energy is preserved to better than $\epsilon_E=\left < \frac{\Delta E}{E_0}\right>\le 0.3\%$ in the axisymmetric models. The magnitude $\epsilon_{divB}=\left<\frac{h~\mathrm{div}~{\mathbf B}}{\vert{\mathbf B}\vert}\right>$, bound to the  divergence constraint ${\mathrm{div}~\mathbf B}=0$, remained  $\epsilon_{divB}\le 2\%$~in all studied cases. 

\subsection{Equivalent resolution and computational effort}
\label{subsec:resolution}

The difference between axisymmetric and full 3D calculations in amount of particles needed to resolve a specific resolution can be highlighted with the concept of \emph{equivalent resolution}. The particle density resulting from homogeneously distributing $N$~particles in a volume $V$ is  $n=N/V$. The inverse of $n$, $v=1/n$, represents the volume of the cell occupied by a single particle. The inter-particle distance is $b = v^{1/D}$, with $D$ being the dimension of the space. Taking $ b$ as the minimum achievable resolution, and assuming that axis-2D and full 3D calculations have equivalent resolutions, i.e.~$ b_{2D}=b_{3D}$, we write:

\begin{equation}
N_{3D}=\frac{V_{3D}}{\left(V_{2D}\right)^{\frac{3}{2}}} N_{2D}^\frac{3}{2}
\label{equivRes1}
\end{equation}

In cylindrical geometry, it is common to consider $V_{2D}= R Z$ and $V_{3D}=\pi R^2 Z$, where $R$ is the radius of the cylinder and $Z$ its altitude. Thus, 

\begin{equation}
    N_{3D}=\pi\left(\frac{R}{Z}\right)^\frac{1}{2} N_{2D}^\frac{3}{2}
    \label{equivres2}
\end{equation}

Many of the calculations reported in this work have $Z=2R$~so that $N_{3D}\simeq 2 N_{2D}^{3/2}$. For a similar spatial resolution, the equivalent number of particles is, in general, much higher in a 3D calculation and, so it is the required computational effort. It is worth to note, however, that some 3D scenarios can be simulated in boxes where one of the sides of the box can be taken smaller than the other two. In these cases, and according to Eq.~(\ref{equivRes1}), any reduction in $V_{3D}$ would significantly reduce the equivalent number of particles $N_{3D}$. A further advantage of axial calculations is that they manage to work with  fewer  neighbor particles, $n_b$, within the kernel range. The default setting is $n_b=60$ and, occasionally $n_b=100$ (the advection loop and cloud collapse tests), which is a factor $\simeq 2-3$ lesser than $n_b \simeq 200-300$  typically used with Wendland interpolators in 3D.         

Additionally, when self-gravity is included in the calculation, the algorithm used to compute the gravitational force also has an impact in the performance of the codes. The ring-like nature of particles in axial symmetric calculations makes it difficult to compute gravity with standard hierarchical methods, such as the Barnes-Hut scheme \citep{barnes86,her89}. As commented in Sect.~\ref{subsec:collapse} gravity can be calculated by computing the direct ring-to-ring interaction \citep{garciasenz2009} which, properly parallelized, is enough to carry out  many   applications with good performance.

In practical terms, we performed a comparison of the average wall-clock time per iteration between our 2D and 3D calculations for two scenarios that will be discussed below: the Z-pinch and the cloud collapse (see Table~\ref{tab:table3}). The former is a pure hydrodynamical simulation, while the latter includes self-gravity. All four simulations were compiled with the same compiler, similar compiler options, and carried out in the same 128-cores AMD Epyc 7742 computing node. As a result the 2D calculations were in average about $\times 33$ faster than the 3D calculation in the Z-pinch case, and about $\times 58$ faster in the collapse case. Such comparisons should, nevertheless, be taken as purely indicative, as we are not only comparing the geometry of the calculation, number particles, and number of neighbors, but also parallelization paradigms, coding infrastructure, memory management, programming languages, and other elements that can speed up or slow down a code considerably. In any case, unless an efficient and scalable algorithm to calculate gravity in 2D-Axial is developed, the advantage of the 2D code will dilute as the number of particles increases when self-gravity is included because of its current scaling order O$(N^2)$. Note, however, that in some    astrophysical scenarios gravity can be handled as arising from a point-like mass, as for example in accretion disks related studies.

\section{Tests}
\label{sec:Tests}

The performance of the axisymmetric formulation is analyzed in light of the comparison between the hydrodynamic code Axis-SPHYNX\footnote{The Axis-SPHYNX code takes advantage of many features of the Cartesian 3D code SPHYNX \cite{cabezon17,gsenz22}, although it does not yet include some sophisticated issues, such as generalized volume elements, grad h terms, or AV switches.} and the well-verified 3D SPMHD code  
 described in \cite{wissing20} which we call GDSPH afterwards, for a suite of test cases. GDSPH is the result of implementing ideal MHD into the code Gasoline2 following the geometric density average scheme. To this end, we have run several MHD standard tests but with fully axisymmetric initial conditions, and  compared the results obtained with both codes for the same initial conditions. As we will see, the match between  Axis-SPHYNX and GDSPH is satisfactory, with minor differences in the results attributable to the initial particle setting, resolution issues, and implementation details.  

The tests that we chose are representative of different physical regimes:

\begin{itemize}

\item Advection and divergence-cleaning: in the advection loop test we aim to explore the robustness of the code to simulating the evolution of magnetized structures on time-scales larger than the characteristic sound-crossing time. It is also a good test to analyze the performance of the divergence-cleaning algorithm (subsect.~\ref{subsec:mloop}).   

\item Explosions: we simulate the evolution of a point-like explosion in a magnetized medium (the magnetic Sedov test). This test is well suited to check the ability of the axisymmetric MHD code to deal with strong shocks (subsect.~\ref{subsec:sedov}).    

\item Implosions: we present the implosion induced by a toroidal magnetic field acting on a plasma sheet moving in an orthogonal direction to it. This test aims to analyze the performance of the code when strong shocks are launched towards the symmetry axis because of the Lorentz-force induced by an azimuthal magnetic field (subsect.~\ref{subsec:Z-pinch}). 

\item Instabilities: we simulated the growth of the Kelvin-Helmholtz instability in a magnetized gas (subsect.~\ref{subsec:KH}). 

\item The collapse of a magnetized and rotating cloud of plasma: this is a rather complete and demanding test which involves many pieces of physics. Besides a barotropic EOS, gravitational and inertial forces have to be incorporated to the numerical scheme (subsect.~\ref{subsec:collapse}).
\end{itemize}

Unless explicitly stated, the default values of the different parameters steering Axis-SPHYNX are those shown in Tables~\ref{tab:table2} and \ref{tab:table3}. The equation of state (EOS) was that of an ideal gas with $\gamma=5/3$ in all tests, except in the collapse of a magnetized cloud in Sect.~\ref{subsec:collapse}, where a barotropic EOS was considered.  Because axial calculations are proner to  suffer from numerical noise and pairing instability, the use of high-order kernels is recommended. By default, Axis-SPHYNX uses the $W_n^s$ sinc family of kernels by \cite{cabezon2008, cabezon17} to carry out interpolations. In particular, we use the $W_5^s$, $W_6^s$ kernels in calculations with $n_b\simeq 60, 100$ neighbors, respectively. The former performing similarly to the quintic, $M_5$ spline. We used the Wendland kernel $C4$ combined with $n_b\simeq 200$ in the GDSPH calculations.

\begin{table}
	\centering
	\caption{Default value of relevant parameters controlling the simulation with Axis-SPHYNX. Columns are:  number of neighbors $n_b$, index $n$ of the sinc kernel $W_n^s$, artificial viscosity coefficients, heat diffusion coefficient ($\alpha_u$) in AV, magnetic dissipation coefficient ($\alpha_B$), and cleaning parameters controlling the $div{\bf B}=0$ constraint. }
 
	\begin{tabular}{lccccccr} 
		\hline
		$n_b$& $n$ ($W_n^s$) &$(\alpha_{AV},\beta_{AV})$&$\alpha_u$&$\alpha_B$&$f_{clean}$&$\sigma_{clean}$& \\
		\hline
		$60~(100)$& $5~(6)$&$(1,2)$&$0.05$&$0.5$&$1$&$1$&\\
		\hline
	\end{tabular}
	\label{tab:table2}
\end{table}

\begin{table}
	\centering
	\caption{ Number of SPH particles ($N$), and minimum value of $h_0$~in the different tests in this work. }
	\begin{tabular}{lccccccr} 
		\hline
		Test & SPH scheme &$N$&$h_0$& \\
		\hline
		Advection Loop&Axis-SPHYNX&$362^2$&$8.0\cdot 10^{-3}$&\\
		Advection Loop&GDSPH&$128^3$&$2.7\cdot 10^{-2}$&\\
		Sedov&Axis-SPHYNX&$362^2$&$8.0\cdot 10^{-3}$&\\
		Sedov&GDSPH&$256^3$&$1.3\cdot 10^{-2}$&\\
		Z-Pinch&Axis-SPHYNX&$362^2$&$8.0\cdot 10^{-3}$&\\
		Z-Pinch&GDSPH&$512^2\times 24$&$6.7\cdot 10^{-3}$&\\
		KH&Axis-SPHYNX&$362^2$&$8.0\cdot 10^{-3}$&\\
		KH&GDSPH&$256^3$&$1.3\cdot 10^{-2}$&\\
		Cloud-Collapse [1]&Axis-SPHYNX&$178^2$&$1.1\cdot 10^{15}$~cm&\\
		Cloud-Collapse [1]&GDSPH&$76^3$&$1.0\cdot10^{16}$~cm&\\
	    Cloud-Collapse [2]&Axis-SPHYNX&$356^2$&$6.0\cdot 10^{14}$~cm&\\
		Cloud-Collapse [2]&GDSPH&$512^3$&$6.1\cdot 10^{14}$~cm&	
		\\
		\hline
	\end{tabular}
	\label{tab:table3}
\end{table}

\subsection{Advection and diffusion of a magnetic loop}
\label{subsec:mloop}

In this test, a weak magnetic loop is advected by a fluid stream moving at constant velocity \citep{gardiner2005,hopkins16}. Grid-based codes have difficulties to describe the evolution of the magnetic loop on many box-crossing periods, owing to the intrinsic dissipation during advection. Nevertheless, the Lagrangian nature of SPH codes makes them ideally suited to this kind of problems and good results for this test have been reported in recent literature \citep{price18,wissing20}.

We consider a cylinder with radius $R=1$~and height $H=2$. The cylinder is filled with an homogeneous, $\rho=1$, flow of plasma moving upwards with uniform velocity $v^z=2$ and constant pressure $P=1$. A spherical magnetic bubble with $R_0=0.3$ and uniform magnetic field ${\mathbf B}= B^{\varphi}
\hat{\mathbf\varphi}$, with $B^\varphi=10^{-3}$, is settled at the center of the cylinder (model $L_A$~in Table~\ref{tab:table4}). Outside of the bubble the magnetic field vanishes, $B^\varphi~(r>R_0)=0$. The spherical magnetic loop is simply advected with the plasma current and nothing is expected to happen. Thus, the loop should keep its initial morphology unchanged during many sound-crossing times. 

Periodic boundary conditions were set on top and bottom of the cylinder, whereas reflective conditions were implemented on its lateral surface and across the symmetry axis. The $v_r$ component of the velocity of particles with $r\simeq 0$ (those with  $r\le h) $ was kept zero during the calculation. With that setting and the initial conditions above, the magnetized bubble periodically returns to the center of the cylinder at times $t=n~T$~($n=1,2,3 ...$), with $T=1$. 
 
\begin{table}
	\centering
	\caption{Relevant magnitudes used in the advection-loop test. Columns are: computed model, initial magnetic vector, adopted value of the cleaning parameter, relative change of the magnetic energy of the loop and normalized value of ${\mathrm{div}~\mathbf B}$~at $t=5~T$. }
	\begin{tabular}{lcccccr} 
		\hline
		Model & ${\mathbf B}_0$&$f_{clean}$ & $\frac{\vert\Delta U_B\vert}{U_{B_0}}$& $\left<\frac{h~\vert\mathrm {div}~{\mathbf B}\vert}{ B}\right>$\\
		\hline
		$L_A$ &$10^{-3}~\hat\varphi$ & $1.0$ & $10^{-2}$ & $0.0$&\\
		$L_B$& $10^{-3}/\sqrt{2}~(\hat z+\hat\varphi)$ & $0.2$ & 0.11 & $9.0~10^{-3}$&\\
		$L_C$&$10^{-3}/\sqrt{2}~(\hat z+\hat\varphi)$   & $1.0$ & 0.16 & $1.4~10^{-3}$&\\
		\hline
	\end{tabular}
	\label{tab:table4}
\end{table}

\begin{figure}
\includegraphics[width=0.50\textwidth]{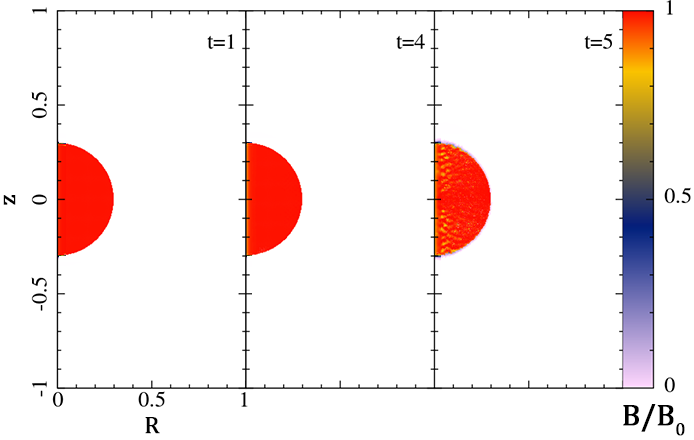}
\caption{Magnetic field strength (normalized to the initial value ${\mathbf B}_0$) of the magnetic loop with ${\mathbf B}= B^\varphi\hat{\mathbf \varphi}$~(model $L_A$~in Table~\ref{tab:table1}), after $t=T$, $4T$, $5T$ complete periods ($T=1$). }
\label{fig:mloop_1}
\end{figure}

Figure~\ref{fig:mloop_1} shows the color map of $\vert {\mathbf B}\vert$~at times $t=T, 4T, 5T$ of model $L_A$ in Table~\ref{tab:table4}. As we can see the magnetic loop preserves its identity until $t\simeq 5T$. At larger times the numerical noise alters the strength of the magnetic field, especially close to the symmetry axis (rightmost panel of Fig.~\ref{fig:mloop_1}). That result is not as good as that in the three-dimensional calculation by \cite{wissing20}, who managed to keep the bubble identity until $t\simeq 20T$, but only a little worse than in \cite{price18} where the bubble evolved neatly well until $t\simeq 5T$. The results with Axis-SPHYNX would improve if a different, more stable, initial distribution of particles is arranged, as for example a Voronoi-like particle setting, which is left for future developments of the code. As commented in Sect.~\ref{sec:boundaries}, particles were spread in a simple square lattice to better handle with the boundary conditions. The smoothing length, $h_b$, is updated at each iteration, so that the number of neighbors per particle it is kept approximately constant around $n_b=100$ in this test.  
 
\begin{figure*}
\includegraphics[width=1\textwidth]{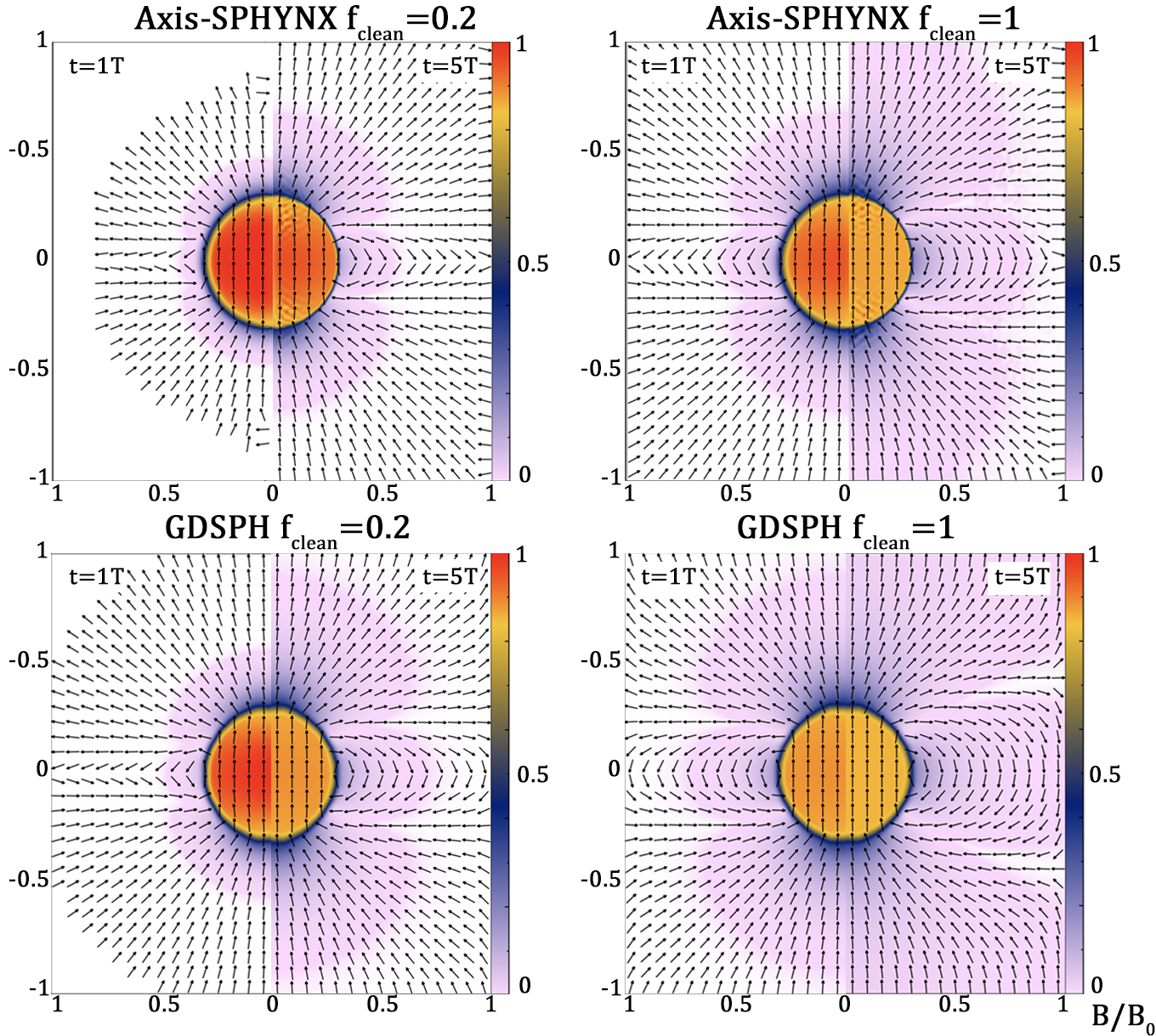}
\caption{Same as Fig.~\ref{fig:mloop_1} but adding a vertical component to the magnetic field, ${\mathbf B}= B^\varphi\hat{\mathbf \varphi}+B^z\hat{\mathbf z}$, which makes $div {\mathbf B}\ne 0$ at the bubble edge (models $L_B$, $L_C$ in Table~\ref{tab:table1}). Upper  panels depict the diffusion of the magnetic field during the divergence cleaning process, as calculated with Axis-SPHYNX with ${\mathrm f_{clean}}=0.2$ and ${\mathrm f_{clean}=1}$, respectively. The same is shown in the lower panels but with GDSPH.}
\label{fig:mloop_2}
\end{figure*}

The first row in Table~\ref{tab:table4} gives more information regarding model $L_A$. The loss of magnetic energy of the magnetic bubble after five cycles, $t=5T$, is rather low, $\simeq 1\%$, with the error in $\mathrm{div}~{\mathbf B}$ being completely negligible. The second and third rows provide the same information as in model $L_A$ but for models $L_B$ and $L_C$, which are not divergence free from the outset. Specifically, in models $L_B, L_C$~the $\hat z$ component of the magnetic field has $\partial{B^z}/\partial {z}\ne 0$~close to the edge of the magnetic bubble. Figure \ref{fig:mloop_2} shows the diffusion of the magnetic field during the process of the $\mathrm{div}~{\mathbf B}$ cleaning at $t=1T, 5T$ and for two values of the cleaning parameter $f_{clean}$. The color maps of $B(r,z)$ obtained with Axis-SPHYNX and GDSPH are qualitatively similar, with the three dimensional calculation showing a bit more diffusion. The geometry of the magnetic field around the bubble, shown by the vector arrows in Fig.~\ref{fig:mloop_2}, is neatly dipolar and  remarkably similar in both calculations.       
 
\begin{figure}
\includegraphics[width=0.50\textwidth]{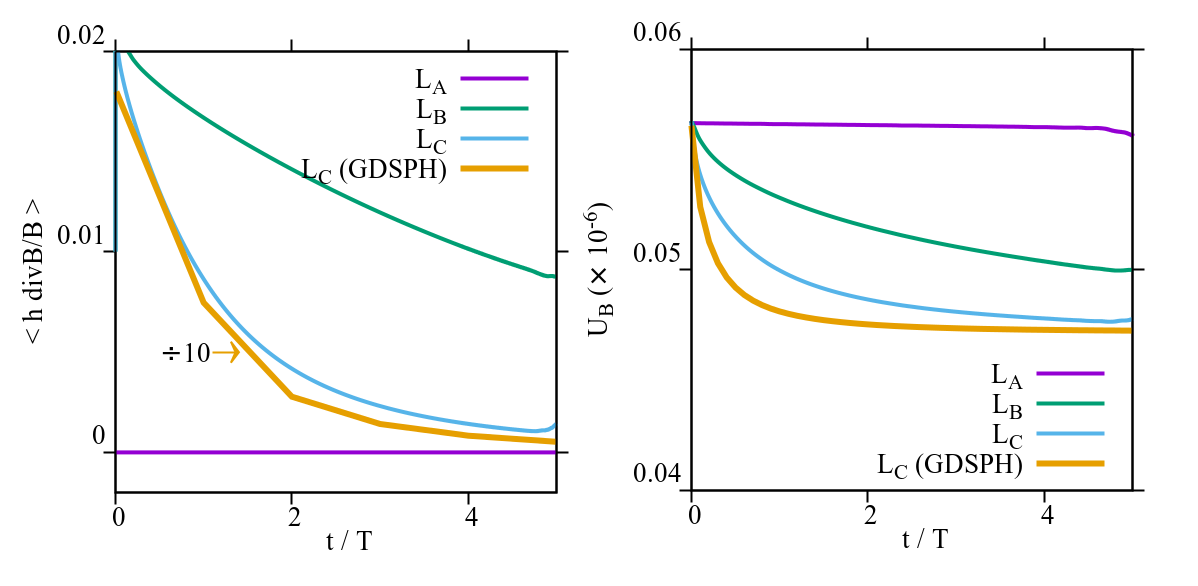}
\caption{Advection loop test. The left panel depicts the evolution of $\mathrm{div}~{\bf B}$ error for models in Table~\ref{tab:table4} until $t=5T$. The right panel shows the magnetic energy content, $U_B$ (in $10^{-6}$ units), of the bubble for the same models.}
\label{fig:mloop_3}
\end{figure}

Figure \ref{fig:mloop_3} depicts the temporal evolution of the magnetic energy in the loop, $U_B$~(in $10^{-6}$ units, right panel), and the magnitude $\left<\frac{h~\vert\mathrm {div}~ {\mathbf B}\vert}{\mathrm{B}}\right>$. As shown in the figure, the evolution of these magnitudes in model $L_A$ is practically flat while the evolution of models $L_B$, $L_C$ strongly depends on the adopted value of the cleaning parameter $f_{clean}$. The default choice, $f_{clean}=1$, gives more diffusion but is much more efficient than $f_{clean}=0.2$~to keep $\mathrm{div}~{\mathbf B}\simeq 0$, as expected.  The magnetic energy content of the loop in model $L_C$ evolves similarly in the GDSPH calculation, although it stabilizes slightly earlier. The absolute value of  $< h~\mathrm{div}~{\bf B}/\mathrm{B} >$ is  almost ten times larger in the 3D  calculation but both, axial and Cartesian, decay fast with similar characteristic times.    

\begin{figure}
\includegraphics[width=0.5\textwidth]{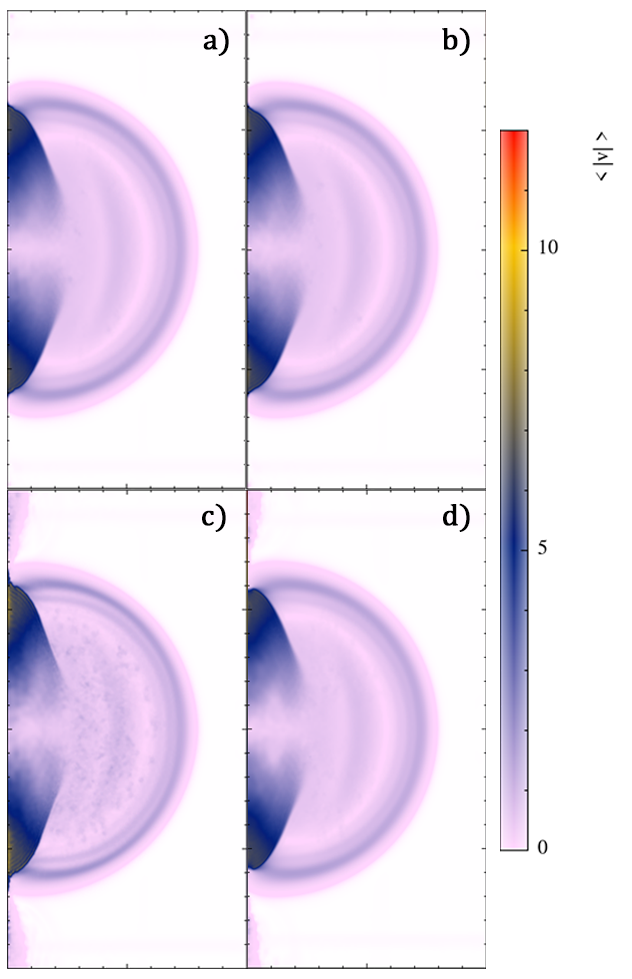}
\caption{ Velocity color-maps at $t=0.048$~ with/without a magnetic noise trigger. The upper-left panel (a) is for the adaptive noise trigger described in the text. The upper-right panel (b) only considers the maximum value of the trigger (i.e. no Balsara corrections). Panel (c) was obtained switching-off the trigger. The lower-right panel (d) is the same as panel (a) but calculated with traditional derivatives instead  the IA.}
\label{fig:sedovnoise_1}
\end{figure}

\begin{figure*}
\includegraphics[width=1\textwidth]{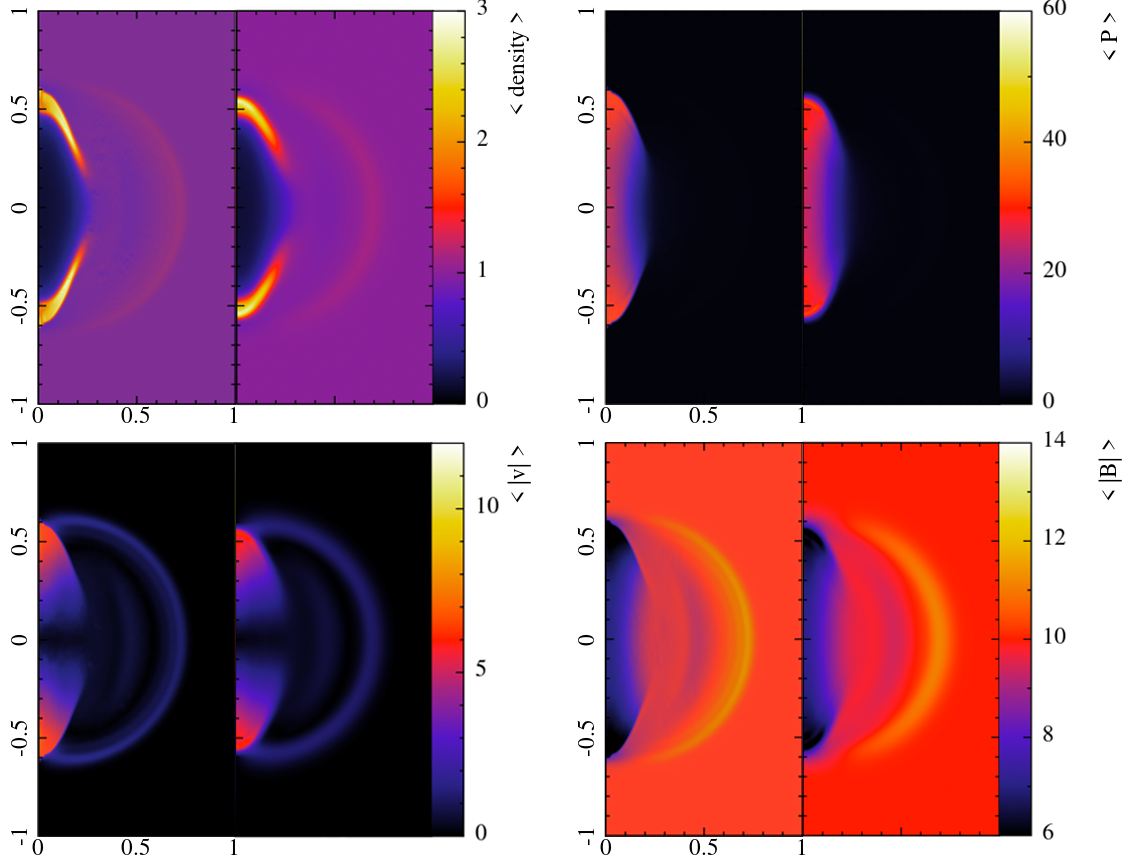}
\caption{Point-like explosion in a magnetized medium calculated with AxisSPHYNX (leftmost sub-figures in each panel) and GDSPH (rightmost sub-figures in each panel.}
\label{fig:sedov_1}
\end{figure*}

\subsection{The magnetic Sedov test}
\label{subsec:sedov}


The axisymmetric version of the MHD Sedov test is easily set by considering an initially spherically symmetric explosion subjected to an external magnetic field $\mathbf B(s,z)= B_z \hat z$. We compare the evolution computed with Axis-SPHYNX to that obtained with GDSPH for the same initial conditions. To seed the explosion we assume a Gaussian initial profile of internal energy: 

\begin{equation}
    u(s)= u_0 \exp[-(s/\delta)^2]+ u_b\,,
    \label{sedov_1}
\end{equation}

\noindent with, 

\begin{equation}
u_0= \frac{E_{tot}}{(\pi^{3/2}~\rho~\delta^3)}\,, 
\end{equation}

\noindent where $E_{tot}=5$ is the total initial energy of the explosion, $\delta=0.1$, and ${\mathbf B} = 10~\hat {\bf z}$. The medium is homogeneous with $\rho_0=1$ and the plasma is an ideal gas with $\gamma=5/3$ and background internal energy $u_b=1$. 
 
An unexpected problem in this test was the large  amount of numerical noise present at late times in the central volume of the box, clearly seen  in Fig.~\ref{fig:sedovnoise_1} (panel c). Such particle disorder is not present in the GDSPH calculation. The noise originates from the feedback between the initial setting of mass points in a lattice and the strong magnetic field. In axial geometry, the initial grid is not as stable as in Cartesian calculations because of the uneven distribution of mass within the kernel range. Even more, any tiny amount of noise in the unshocked region is magnified by the magnetic field. Unfortunately, raising  the number of neighbors without increasing the total number of particles did not solve this issue.

To face this problem we introduce a magnetic noise-trigger, $\mathcal{N}\mathcal{T}_B$, which  keeps the artificial viscosity sufficiently high to counter-balance the residual magnetic force in the unshocked region. In our case, it is sufficient to steer the Balsara limiters, with $\zeta=\frac{B^2}{2\mu_0~P}$, where $\zeta$ is the the inverse of the $\beta$-plasma parameter: 

\begin{equation}
\mathcal{N}\mathcal{T}_B : f_a=
\begin{cases}
1 &\text{if } \zeta \ge 0.5 \,, \\
max[f^0_a, \frac{2}{3}\left(5\zeta-1\right)] &\text{if } 0.2 <\zeta < 0.5\,,\\
f^0_a &\text{if } \zeta\le 0.2\,,
\end{cases}
\label{triggerscheme}
\end{equation}

\noindent
 where $f^0_a$~is the limiter given by Eq.~(\ref{balsara}) and $f_a$~is the final adopted value.  The impact of including or not $\mathcal{N}\mathcal{T}_B$ in the simulations is shown in Fig.~\ref{fig:sedovnoise_1}, which depicts the color map of velocity at $t=0.048$ in four cases. These have been calculated with $\mathcal{N}\mathcal{T}_B$ switched on (panel a), off (panel c) and without Balsara corrections (panel b), whereas panel d was obtained with traditional derivatives and $\mathcal{N}\mathcal{T}_B$ switched on. The results convincingly show that the inclusion of a magnetic noise trigger significantly reduces the particle disorder. For this problem, calculating  the derivatives with the IA or with the  traditional scheme leads to similar results,  but the latter is slightly noisier. 

Figure~\ref{fig:sedov_1} summarizes the results of the calculations and  Table~\ref{tab:table3} shows additional information regarding the total number of particles used in this test and on the resolution. Simulations with Axis-SPHYNX make use of the magnetic noise-trigger,  Eq.~(\ref{triggerscheme}), to keep the system more ordered before the arrival of the shock wave. The color maps depicting the density, pressure, modulus of velocity, and magnetic field at $t=0.048$, do not show significant differences between the simulations carried out with Axis-SPHYNX (leftmost sub-figures) and GDSPH (rightmost sub-figures). They   qualitatively agree with the results published by \cite{rosswog2007}, who simulated a similar explosion but inside a weaker magnetic field. The shock front is slightly ahead in the axisymmetric case, which is due to the higher resolution in that calculation. The color maps of the velocity modulus (third panels) show extended regions with low velocity at $r\le 0.5$, well captured  in both cases.  
The relative error of total energy is low, $\epsilon_E\simeq 10^{-2}~\%$, at all times. The estimator $\epsilon_{div{\mathbf  B}}$ measuring the averaged deviation of the constraint $\mathrm{div}~{\mathbf B}$ was always $\epsilon_{div{\mathbf B}}\le 0.2\%$.


\subsection{Z-pinch like implosion}
\label{subsec:Z-pinch}

\begin{figure*}
\includegraphics[width=1\textwidth]{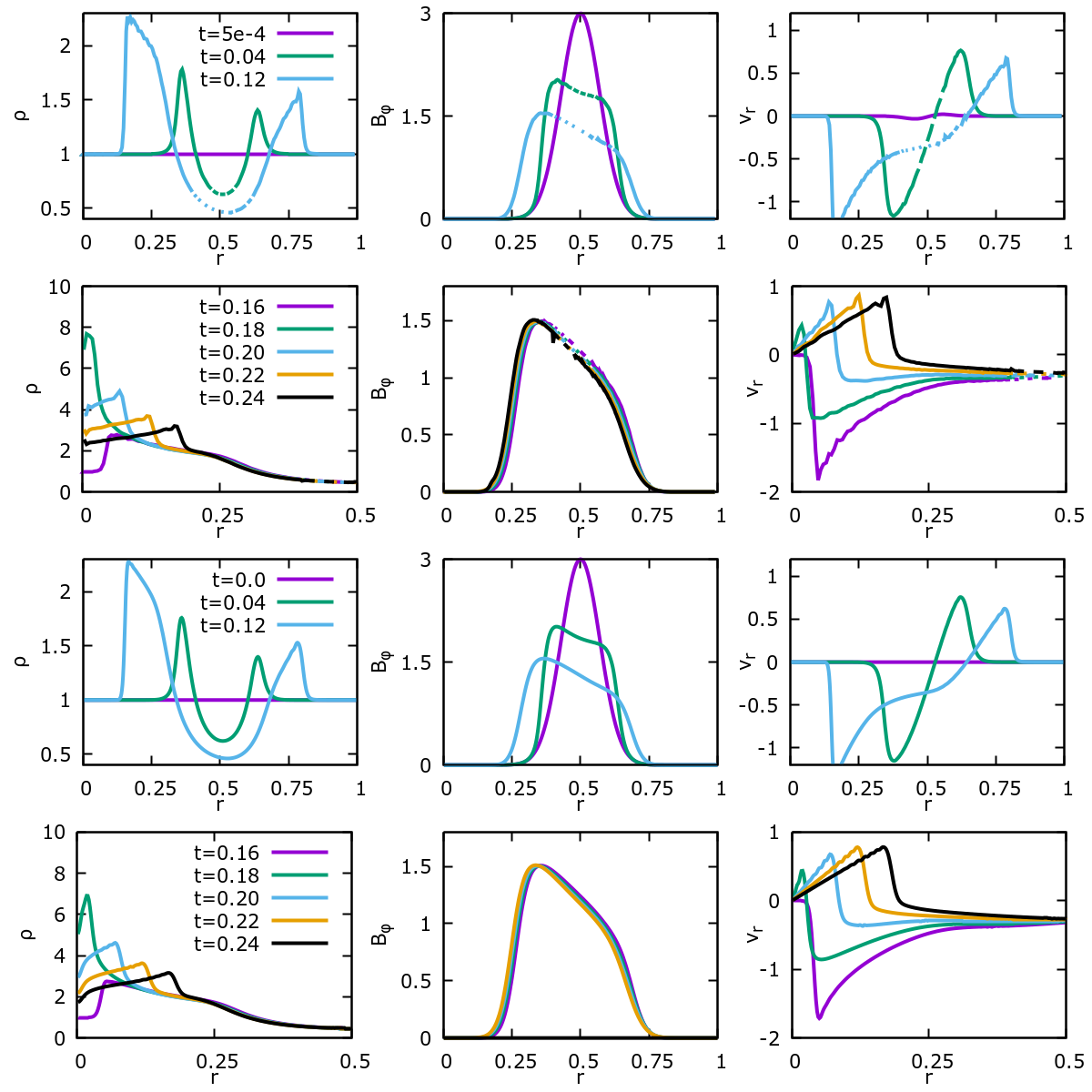}
\caption{Implosion in a magnetized medium (Z-pinch test) calculated with Axis-SPHYNX (first and second rows) and with GDSPH (third and fourth rows). The figure shows the shell-averaged values of density ($\rho$), azimuthal component of the magnetic field ($B_\varphi$), and radial velocity ($v_r$).}
\label{fig:zpinch_1}
\end{figure*}

The so-called Z-pinch devices were among the first to explore the feasibility of having controlled nuclear fusion on Earth \citep[for a review]{haines2000,shumlak20}. They have also been applied to conduct many laboratory astrophysics experiments \citep{ciardi2004,Bocchi13}. In the Z-pinch machines a strong toroidal magnetic field, $B^\varphi$ is created by a mega-amp\`ere electric current pulse ($\simeq 1\mu s$) moving in the axial direction. The Lorentz force exerted by $B^\varphi$ on the plasma, which initially moves in the Z-direction, impels it towards the Z-axis. The compression of the plasma at the symmetry axis can be strong, provided that the initial conditions have a good degree of axial symmetry.  

To sketch the basic physics of a magnetic Z-pinch process in a simple numerical experiment, we consider an initially homogeneous plasma with $\rho=1$, $P=1$ in a cylinder with radius $R=1$ and height $Z=2$. The plasma is initially moving with $v_z= -1$. A toroidal magnetic field, $B^\varphi$, with maximum strength $B_0=3$ and with a gaussian profile,  

\begin{equation}
    B^\varphi= B_0~\exp\left[-(r-r_0)^2/\delta\right]\,,
    \label{zpinch_1}
\end{equation}

\noindent which is set at around coordinate $r_0=0.5$ with characteristic width $\delta=0.01$. The boundary conditions are periodic on top and bottom of the cylinder and reflective on the lateral surface. As in the point like explosion test, we aim to compare the results with Axis-SPHYNX to those obtained with  GDSPH and identical initial conditions. Table~\ref{tab:table3} shows the number of SPH particles and initial resolution, $h_0$. 

We present the main results of this numerical experiment in Fig.~\ref{fig:zpinch_1}. That figure depicts the profiles of the r-averaged magnitudes of $\rho$, $B^\varphi$, $v^r$, at different elapsed times. The first and second rows correspond to the axisymmetric calculation, whereas the other two resulted from the full three-dimensional calculation with GDSPH. As we can see, the match between the results obtained with both codes is excellent. The main difference is that the density peak around the point of maximum compression at $t=0.18$ is a slightly larger in the axisymmetric calculation. The toroidal component of the magnetic field evolves very similarly in both simulations. The profile of the radial velocity is particularly sensitive to the magnetic part in the hoop-stress term in Eq.~(\ref{mhdaccel_r}). Nevertheless, the $v^r$ profile at the supersonic shock front is sharp and well captured by both codes. The evolution after the rebound, $t\ge 0.18$, is also very similar. The profiles of $v^r$  obtained with Axis-SPHYNX are not as smooth as those calculated with GDSPH, owing to the lesser number of neighbors used to carry out the interpolations ($n_b\simeq 60$ in the former and $n_b\simeq 200$ in the latter). In this test, the total energy was preserved up to $\frac{\Delta E}{E_0}\le 0.4\%$ and the constraint $\mathrm{div}~{\mathbf B}=0$ was fulfilled to machine precision. 

	

\subsection{Magnetic Kelvin-Helmholtz instability}
\label{subsec:KH}

\begin{figure*}
\includegraphics[width=1\textwidth]{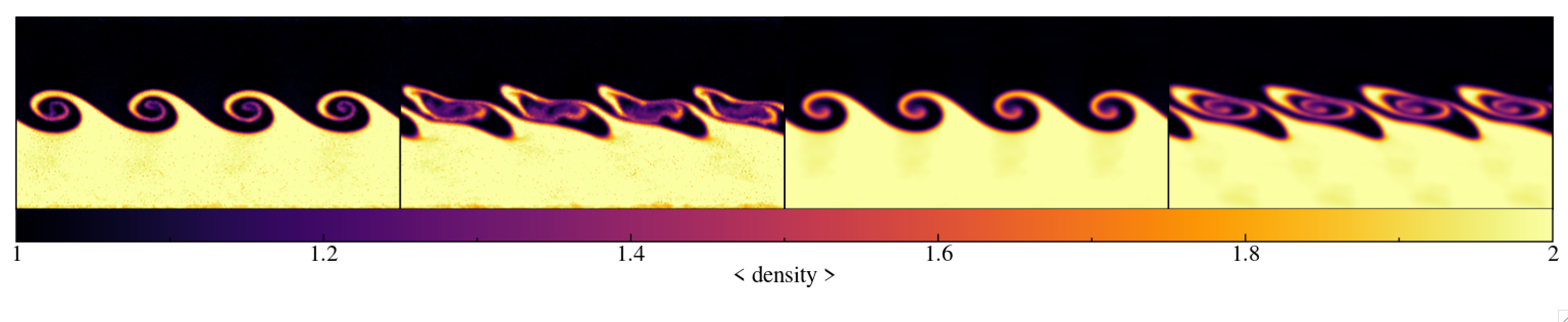}
\caption{Particles distribution in the KH experiment at two elapsed times, $t=1.5$~and $t=2.8$, for models calculated with AxisSPHYNX (first two panels) and  calculated with GDSPH (XZ slices in the two rightmost panels).}
\label{fig:kh_1}
\end{figure*}

The growth of the Kelvin-Helmholtz instability across the contact layer between fluids with different densities is a challenging test for hydrodynamic codes. Resolution issues limit the growth rate of the instability during the initial linear stage to, later on, hinder the development of small wave-lengths in the non-linear phase \citep{mcn12}. Modern SPH codes are able to cope with the KH instability, even when a relatively low number of particles is used \citep{rosswog20}, but provided  the density contrast is not very large. Adding a magnetic field to the plasma turns this test into an interesting, albeit more complex, MHD problem where we expect some alignment of the billows with the dominant direction of the magnetic field. 

\begin{figure}
\includegraphics[width=0.48\textwidth]{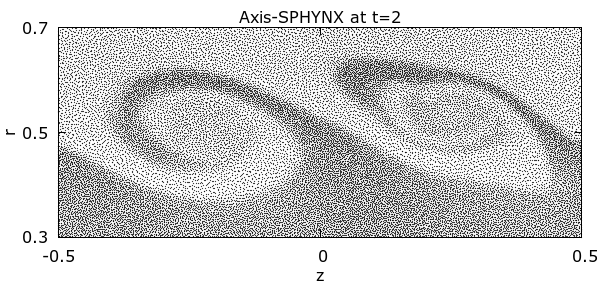}
\caption{ Zoom showing the distribution of particles at t=2 as calculated with Axis-SPHYNX. The fluid inter-phase is smooth and free of gaps, with no indications of the tensile instability.}
\label{fig:kh_2}
\end{figure}

Three-dimensional SPH simulations of the growth of the KH instability in a weakly magnetized medium have been reported by \cite{hopkins16,wissing20}. The main effect of the magnetic field is to uncoil and stretch the vortexes during the non-linear stage, so that the instability looks rather different from that of non-magnetized systems. The axisymmetric realization of these 3D-MHD experiments is similar to that described in the papers above. It consists on two interacting fluids  moving along two concentric cylindrical pipes, but in opposite directions. An uniform magnetic field, $B^z$, pointing along the axis of the pipe, is added so that it interacts with the radial component of the velocity, $v_r$, of the unstable layer via the Lorentz-force.

\begin{figure*}
\includegraphics[width=1\textwidth]{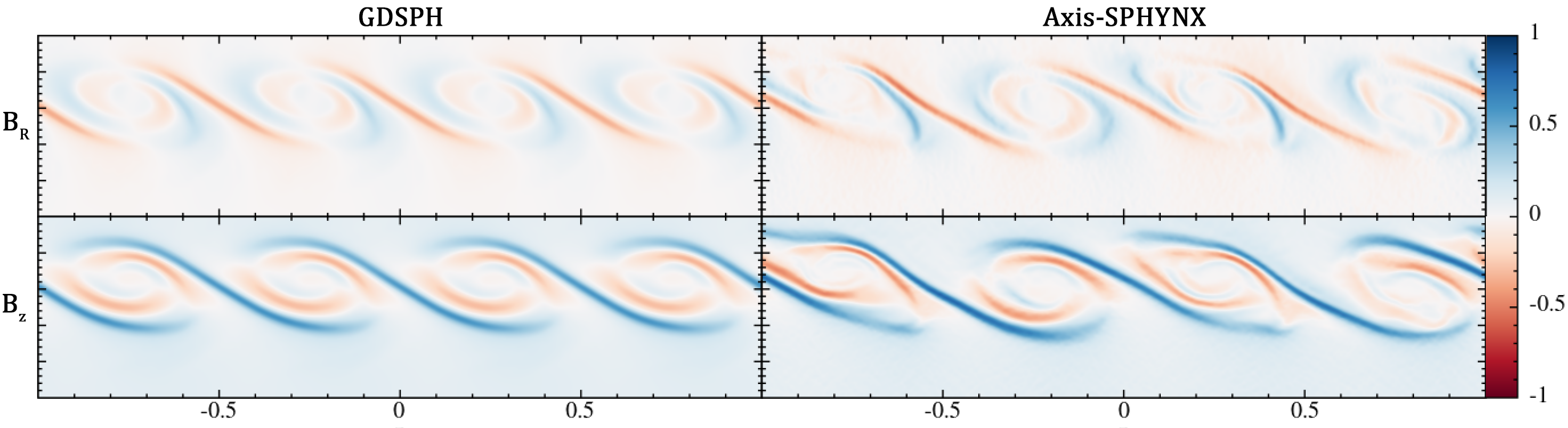}
\caption{Color-map depicting the distribution of the radial (upper panels) and axial (lower panels) components of the magnetic field at $t=2$~in the KH simulation.}
\label{fig:KH_B}
\end{figure*}

We consider a cylinder with radius $R=1$ and longitude $L=2$. A fluid with density $\rho_{in}=2$ moving with $v_z=+0.5$ fills the inner half, $r\le R/2$, of the cylinder. The outer part of the cylinder is filled with a lighter fluid, $\rho_{out}=1$, moving with $v_z=-0.5$. Both fluids share the same pressure, $P=2.5$, and are immersed in a magnetic field $B_z=0.1$. The inner and outer fluids are simulated  with two square lattices with sizes according to the density contrast. The fluid interface was not smoothed, and it  was altered by adding a small radial perturbation to  $v_r$,        

\begin{equation}
   v^r=\Delta v^r\exp\left(-\frac{\vert r-0.5\vert}{0.1}\right)\sin\left(4\pi z\right)
   \label{kh_2}
   \end{equation}
   
\noindent with $\Delta v^r=0.05$. Table~\ref{tab:table3} shows the total number of particles and initial maximum resolution. 

Figure~\ref{fig:kh_1} depicts the color-map of the density at two times, $t=1.5$ and $t=2.8$ which are representative of the early and evolved non-linear phase respectively\footnote{ The characteristic growth-time in the plane-parallel approximation can be taken as a rough reference, $\tau_{KH}\simeq 1.08$}. The density maps at $t=1.5$ are rather similar, with the axisymmetric calculation showing more structure owing to the higher resolution and more sensitive estimation of gradients.  During the advanced non-linear stage, $t=2.8$ in Fig.~\ref{fig:kh_1}, the cumulative effect of the magnetic force stretches the vortex along the symmetry axis of the cylinder and the morphology of the billows differ. The axial calculation shows more distorted billows than the Cartesian simulation and with less rounded tips. This is due to the different sensitivity of the numerical schemes  used to compute the gradients. The integral approach is more sensitive to the initial setting of particles in two square grids of different size around the interface.  Smaller initial asymmetries grow more efficiently during the non-linear  stage in the axial calculation. A way exert control on such sensitivity was to  raise the floor value of the Balsara limiters from its default setting, $f_a^{floor}=0.05$ to $f_a^{floor}=0.3$ to better retain the identity of the billows \footnote{ A plot depicting the evolution during the non-linear phase with $f_a^{floor}=0.05$, showing more asymmetrical billows at $t=2.8$ can be found in  \cite{gsenz2022}}.


A zoom of the particle distribution around the two central billows at $t=2$ is shown in Fig.~\ref{fig:kh_2}. The contact surface between the dense and light fluids is clean and continuous, showing no gaps or any trace of the tensile instability. 


The suitability of using a Lagrangian method to describe instabilities in presence of magnetic fields is further highlighted in Fig.~\ref{fig:KH_B}, which depicts the geometry of $B^r$ (upper panels) and $B^z$ (lower panels) at $t=2$. The magnetic field is well threaded along the distorted plasma stream-lines and vortexes, with the radial and axial components drawing lines in phase opposition. There is a good agreement between the axial and the three-dimensional calculation. 

	

\subsection{Collapse of a rotating-magnetized cloud}
\label{subsec:collapse}

The collapse of a rotating and magnetized dense cloud of gas embedded in a more dilute medium has become a standard test to verify MHD hydrodynamic codes \citep{hennebell2008,hopkins16}.  Outflows from gravitationally collapsing magnetized dense gas clouds were obtained for first time with SPH by \cite{pri12b}. This test involves many physical ingredients of astrophysical interest such as gravity, rotation, and magnetic fields. Because the collapse of the cloud basically proceeds with axial geometry (except in those cases where there is fragmentation), this scenario can be approached with axisymmetric MHD codes. 

The initial setting is the same as in \cite{wissing20}. A cloud with mass $M=1~M_{\odot}$ and density $\rho_C= 4.8\cdot 10^{-18}$\dens, rotates around the Z-axis with $\omega_0= 4.24\cdot 10^{-13}$~s$^{-1}$. The cloud is surrounded by background interstellar medium (ISM), with a radius ten times larger  and density $\rho_{ISM}=\rho_C/300$. The whole system is inside a magnetic field $\mathbf B = \frac{610}{\mu} \hat z~\mu G$ aligned with the rotation axis of the cloud, where $\mu$~is a parameter steering the intensity of the magnetic field. Three-dimensional simulations of the collapse, with a barotropic EOS, 

\begin{equation}
P=c_{s,0}^2\sqrt{1+\left(\frac{\rho}{\rho_0}\right)^{\frac{4}{3}}}\,,
\label{Pbarotropic}
\end{equation}

\noindent with $\rho_0=10^{-14}$~\dens and $c_{s,0}=0.2$~km$\cdot s^{-1}$, have shown that the implosion of the cloud would produce a narrow jet only if the parameter $\mu$ is neither too large, nor too small: $2\le \mu \le 75$ \citep{hopkins16, wissing20}. 

This test is challenging for an axisymmetric SPH code because the collapse is strong and impels the particles towards the singularity axis. The central density increases five orders of magnitude and the Courant criterion enforces the time-step to be extremely small. In this test, we want to check if Axis-SPHYNX is able to reproduce the main features of the collapse of the cloud, as for example the maximum achieved density, the equatorial flattening of the cloud, and the jet emergence at around the free-fall time of the cloud, $t_{ff}=\sqrt{\frac{3}{2\pi G\rho_C}}\simeq 1.2\cdot 10^{12}$~s. We carried out three simulations of this scenario with $\mu=\infty, \mu=20, \mu=10$, from the initially spherically symmetric conditions until the formation of the disk, and the beginning of the jet launch at $t\simeq 1.1\cdot 10^{12}$~s. 

The gravitational force, ${\mathbf g}$, is calculated using the scheme described in \cite{garciasenz2009} and is added to the acceleration. Basically, self-gravity is calculated with direct ring-to-ring interactions, first computing the gravitational potential, $V_g$, to later make the SPH estimation of its gradient ${\mathbf g}=-{\mathbf \nabla V_g}$. Obviously, this results in a larger computational effort than in the previous, gravity-free tests but it is still lower than that invested by GDSPH for the same scenario, owing to the large differences in the number of particles in both codes (Table~\ref{tab:table3}) and average number of neighbors ($\times~2$~ in GDSPH). 

For this problem, it is better to calculate the specific angular momentum $\ell^{z} = r v^\varphi$,  rather than $v^\varphi$, so that in absence of azimuthal forces the angular momentum is conserved. The momentum equations, Eqs.~(\ref{mhdaccel_r}), (\ref{mhdaccel_z}), and (\ref{mhdaccel_phi}), become, 

\begin{equation}
    \frac{dv^r_a}{dt}= a^r_a + g^r_a + \frac{(\ell^z)^2}{r^3}\,.
    \label{cloud_r}
\end{equation}

\begin{equation}
    \frac{dv^z_a}{dt}= a^z_a + g^z_a\,.
    \label{cloud_z}
\end{equation}

\begin{equation}
    \frac{1}{r}\frac{d \ell^z}{dt}=a^\varphi\,.
    \label{cloud_phi}
\end{equation}

\begin{figure}
\includegraphics[width=0.50\textwidth]{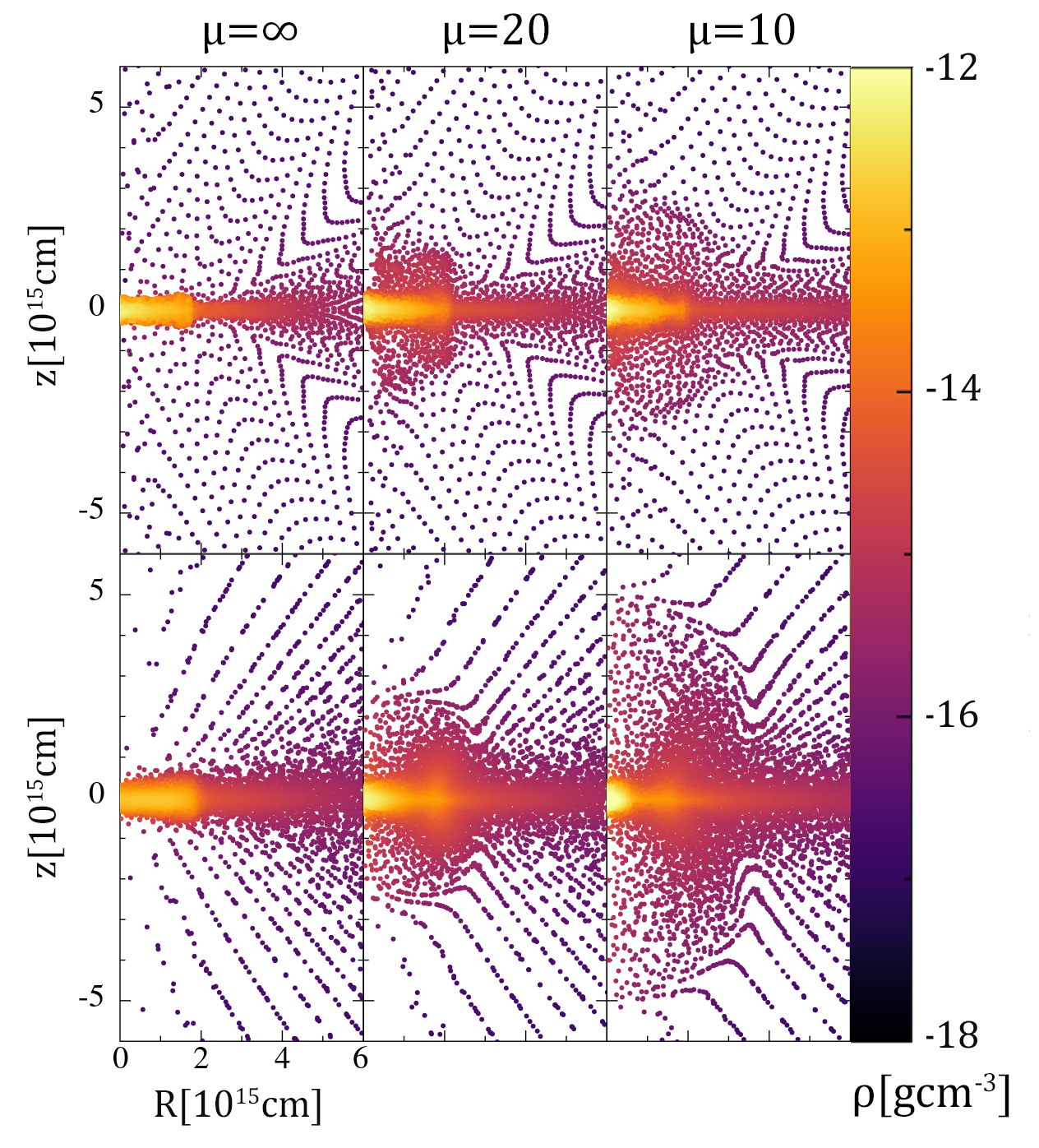}
\caption{Density color-maps of the core  of the collapsing cloud at common elapsed time $t=1.1\cdot 10^{12}$~s for the low resolution calculations (Cloud Collapse [1] in  Table~\ref{tab:table3}). The upper panels show the results with Axis-SPHYNX for three values of the magnetic field,  $B^z=610/\mu$. The same is shown in the lower panels, but calculated with the code GDSPH.}
\label{fig:cloud1}
\end{figure}

\begin{figure}
\includegraphics[width=0.50\textwidth]{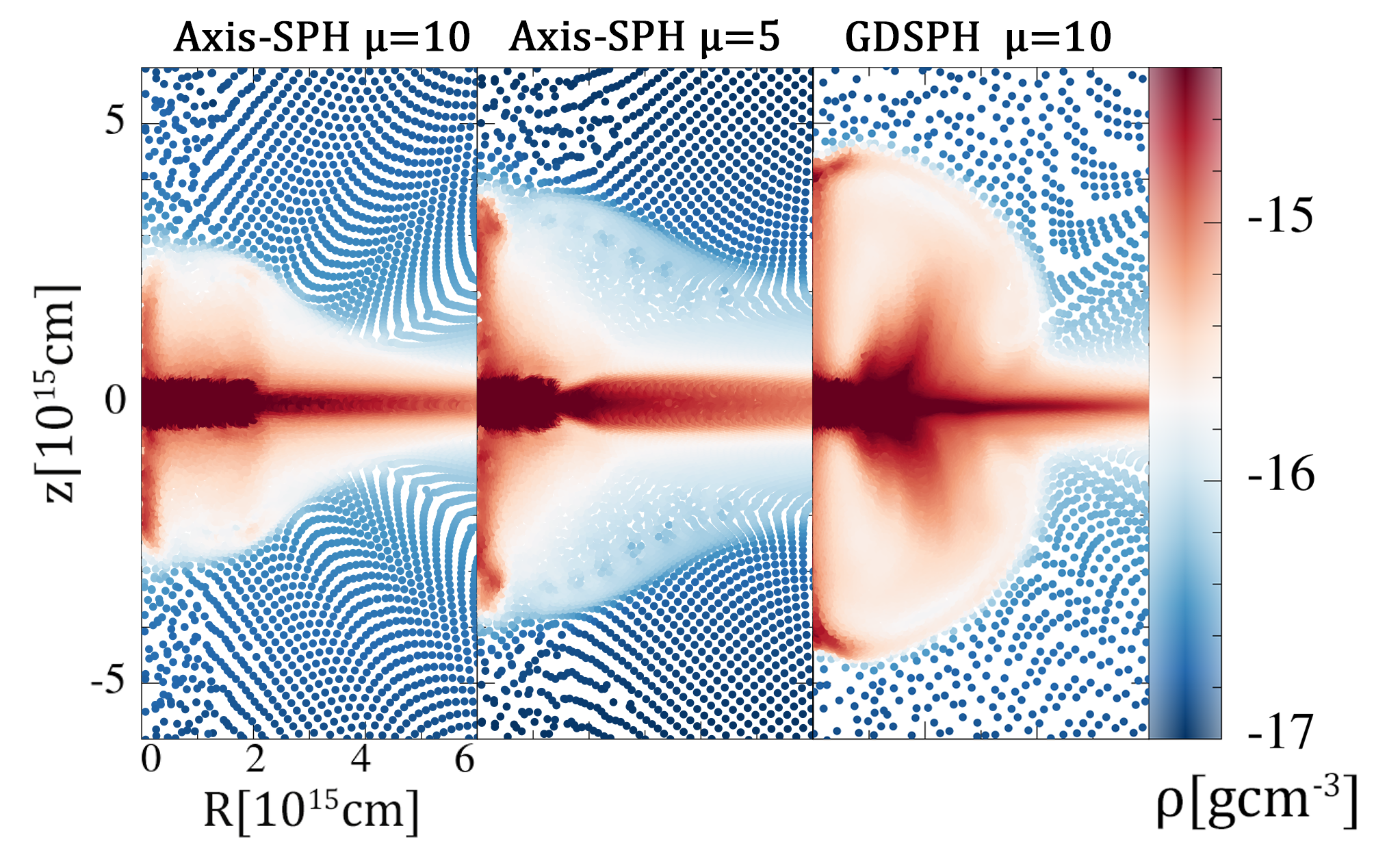}
\caption{Emergence of collimated jets at the core of the collapsing cloud at elapsed time  $t=1.1 \cdot 10^{12}$~s and $\mu=10, \mu=5$, obtained with enhanced resolution (Cloud Collapse [2] in Table~\ref{tab:table3}). The last  panel shows the results with GDSPH at the same elapsed time and $\mu=10$, and depicting a slice cut in plane XZ.}
\label{fig:cloud2}
\end{figure}

\begin{figure}
\includegraphics[width=0.48\textwidth]{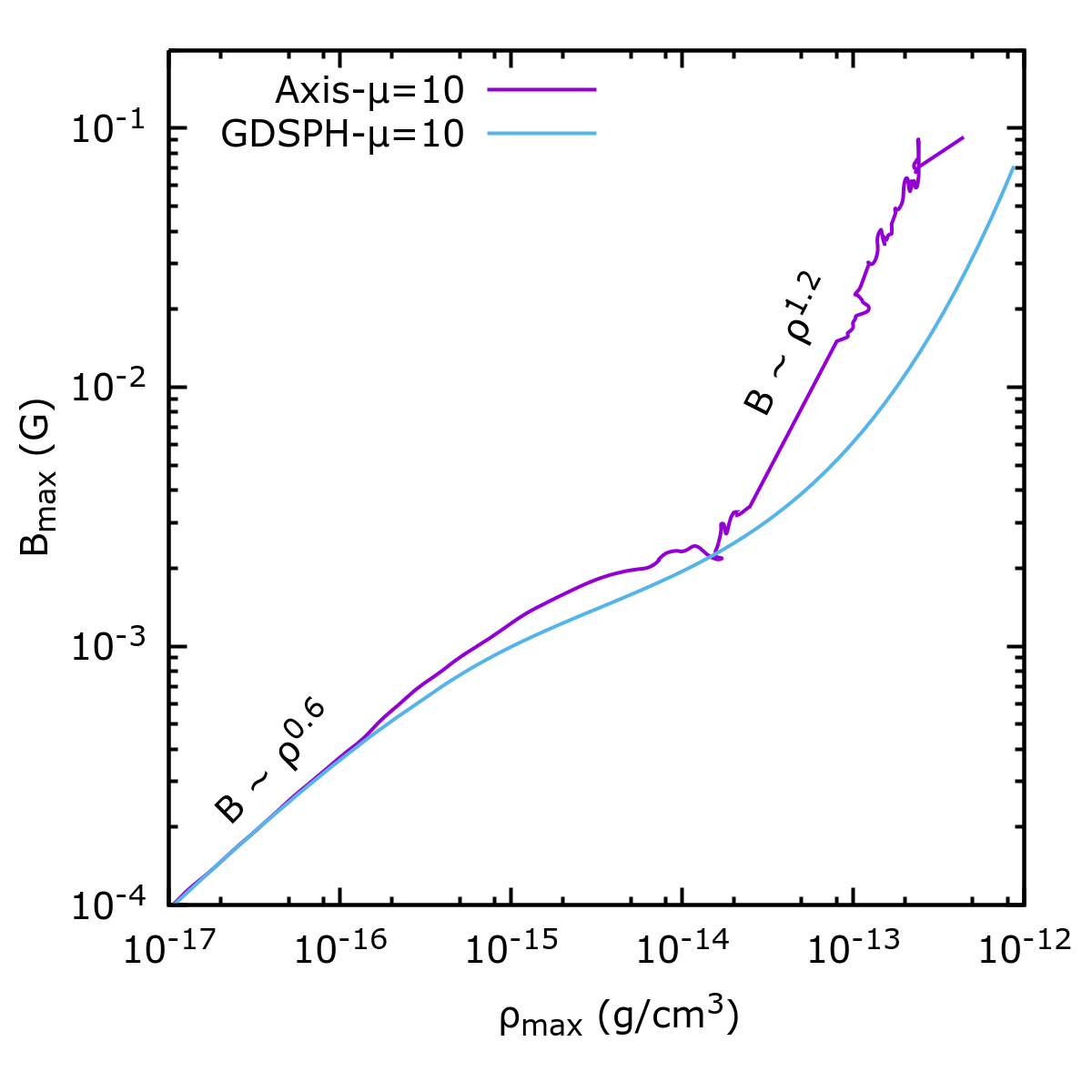}
\caption{Maximum magnetic field $B_{max}$~versus maximum $\rho_{max}$ for the case $\mu=10$ calculated with Axis-SPHYNX (magenta line) and GDSPH (blue line). }
\label{fig:cloud3}
\end{figure}

Figure~\ref{fig:cloud1} shows the density color map of the innermost region of the cloud at $t=1.1\cdot 10^{12}$~s, when the jets, if any, are born. That time is close to the free-fall time $t_{ff}\simeq 1.2\cdot 10^{12}$~s. The upper row of panels depict the calculation with Axis-SPHYNX and the lower row is for the GDSPH calculation, both evolved from the initial models with lower resolution in Table~\ref{tab:table3}. The color map from the GDSPH calculation was built taking a slice in cylindrical coordinates with width $\Delta \varphi=0.05$~rads. Both panels look similar, with  the axisymmetric calculation being a bit less evolved than its 3D counterpart. At $t=1.1\cdot 10^{12}$~s the cloud has already collapsed into a disk with similar central densities, $\simeq 10^{-12}$\dens, in both cases. The two codes indicate the same qualitative trend with decreasing values of the $\mu$ parameter. A high value, $\mu\rightarrow\infty$~(i.e. $B^z\simeq 0)$ there is no jet at all, whereas suitable conditions for jet formation are seen for $\mu < 20$, which is encouraging. Nevertheless, no collimated jets attached to the rotation axis were observed in these low-resolution calculations with Axis-SPHYNX. Simulations with enhanced resolution follow a similar trend, but this time clear collimated jets develop, especially with $\mu < 10$. This is shown in Fig.~\ref{fig:cloud2} which emphasizes the color intensity around the axis region so that the jets are better highlighted. As it can be seen, the calculation with $\mu=10$ gives birth to a jet, albeit lesser developed than its 3D counterpart (last snapshot in Fig.~\ref{fig:cloud2}). On another note  choosing $\mu=5$ produces a more robust and well-developed jet. Although these results suggest that for this kind of problems some improvement of the axial calculation is still necessary, the outcome is roughly consistent with the calculations by \cite{hopkins16} who did not find collimated jets in their low-resolution SPH calculations, needing from low $\mu$-values ($\mu<10$) to observe well-developed jets with higher resolution. 

Several causes can contribute to the difficulties in building the jet in the Axis-SPHYNX calculation and to the observed dissimilitudes at  $t\simeq t_{ff}$. The first  could be attributable to the slightly different representation of the azimuthal velocity field at the very center of the collapsed cloud between the axial and the 3D simulations. In three-dimensional calculations the  velocity $v^\varphi$ is smoothed by the AV close to the center of the disk which approaches rigid rotation. In axial geometry, however, the AV only works in the $\{r, z\}$~plane and $v^{\varphi}$~is not smoothed. A plausible remedy to that behavior is to add some amount of shear viscosity to $a^\varphi$, with the scheme developed by \cite{Sijacki2006} for example, which is left to future extensions of the code. It could also be possible that the mixing of particles with different masses during the anisotropic collapse leads to a build up of the numerical errors in the central region. In this respect, a possibility worth to explore is to consider SPH formulations which are less dependent on the mass of the particles \citep{ott2003}.


 The follow-up of the maximum strength of the magnetic field, $B_{max}$ versus the maximum density, $\rho_{max}$ is a good indicator of the collapse process \citep{Wurster2018}. Figure \ref{fig:cloud3} shows the profile of $B_{max}$ as a function of the maximum density at common elapsed times for $\mu=10$. The match between the 2D-axial and the 3D calculation is pretty good until $\rho_{max}\simeq 3\cdot 10^{-14}$ \dens, when non-linear effects take over. The agreement is qualitative hereafter. But both calculations follow the same trend, showing a similar large increase of the slope of the profile at $\rho_{max} > 3\cdot 10^{-14}$\dens.   Near $\rho_{max}\simeq 10^{-12}$\dens there is a factor 3-5 difference between both calculations, which is not a surprise given the sensitivity of $B_{max}$ on implementation details, as for example the amount of magnetic dissipation. Such strong dependence of the $B_{max} (\rho_{max})$ trajectory on implementation details (Ohmic and ambipolar diffusion, Hall effect), was also reported by \cite{Tsukamoto2015} and \cite{Wurster2018} although with  different initial conditions, EOS and in simulations spanning a wider density range.


\section{Conclusions}

In this work, we propose a novel SPH formulation of ideal magneto-hydrodynamics with axial geometry and provide the basic pieces to build an axisymmetric SPMHD simulation code. The main goal is to tackle problems with higher resolution and lower computational effort than standard SPMHD codes. Such computational tool can be of interest not only to astrophysicists but to plasma researchers in general.  The proposed scheme and its associated hydrodynamic code, called Axis-SPHYNX, have been verified by direct comparison with the results of the three-dimensional SPMHD code  by \cite{wissing20}. 

On the whole, there is a good match between both hydrodynamic codes in the performed tests, with the axial approach showing a bit more numerical noise, especially close to the symmetry axis.  Axisymmetric SPH calculations are intrinsically noisier than Cartesian, owing to the uneven distribution of mass within the kernel range, even in homogeneous systems. Furthermore, they are more prone to undergo pairing instability and the use of high-order interpolators is recommended. In calculations involving low plasma-$\beta$ values (i.e. in the strong field regime) the use of a magnetic noise-trigger, such as that in Eq.~(\ref{triggerscheme}), helps to prevent the growth of the numerical  noise. Looking for both, more stable initial models and procedures to control particle disorder deserve future work. The agreement with GDSPH is excellent in the case of simulating explosions and implosions in magnetized systems, which could be of interest to understanding the physics of plasma compression in terrestrial laboratories. The axisymmetric code is also able to simulate the growth of instabilities in magnetized plasmas, such as the Kelvin-Helmholtz instability, which involves longer time-scales than explosions. Even though the axial formulation of SPMHD does not guarantee complete conservation properties,  we found that energy and momentum in the Z-direction (not affected by hoop-stress forces) are preserved  $\le 0.1\%$. The averaged divergence constraint $\left<  h~\mathrm{div}~ \mathbf{B}/\mathrm{B}\right>$~remained below $2\%$~in all the tests. 

Axis-SPHYNX can handle more complex scenarios such as those involving gravity and rotation, of indisputable interest to astrophysics. As shown in Sect.~\ref{subsec:collapse}, with the collapse of a magnetized cloud, the proposed scheme is able to successfully cope with that scenario. There is a quantitative agreement between the two codes during the nearly free-fall phase of the collapse and further formation of the high-density, rotating disk at the equator. Nevertheless, in more advanced stages the agreement between both codes is basically qualitative and work has to be done to enhance the calculations. For example, one should consider the role of the shear viscosity in the evolution of the azimuthal component of the velocity, $v^\varphi$. Immediate prospects are to incorporate grad-$h$ effects, AV switches, as well as to improve the initial model generation, and to refine the treatment of particles that move close to the singularity axis.  

\section*{Acknowledgements}

We thank the anonymous referee for the useful comments and suggestions which have improved the quality and presentation of the paper. This work has been supported by the MINECO Spanish project  PID2020-117252GB-100 (D.G.), by the Swiss Platform for Advanced Scientific Computing (PASC) project SPH-EXA: Optimizing Smooth Particle Hydrodynamics for Exascale Computing (R.C. and D.G.). The authors acknowledge the support of sciCORE (http://scicore.unibas.ch/) scientific computing core facility at University of Basel, where part of these calculations were performed. The GDSPH simulations were performed using the resources from the National Infrastructure for High Performance Computing and Data Storage in Norway, UNINETT Sigma2, allocated to Project NN9477K. We also acknowledge the support from the Research Council of Norway through NFR Young Research Talents Grant 276043. Author M.L. received support from the European Research Council (ERC) under the European Union's Horizon 2020 research and innovation programme (grant agreement No. 101002352). 

 The color map figures were drawn with the visualization tool SPLASH  \citep{price2007}.

\section*{DATA AVAILABILITY}

The data supporting this article will be shared on reasonable request to the corresponding author. A current version of Axis-SPHYNX is publicly available at:~\href{https://github.com/realnewton/AxisSPHYNX}{Axis-SPHYNX download}




\bibliographystyle{elsarticle-harv}




\appendix

\section{Derivation of the axisymmetric SPH equations}
\label{sec:appendix A}

A simple procedure to obtain different flavors of the SPH equations of momentum and energy is to consider the following identity \citep{rea10}, 

\begin{equation}
    \frac{d\mathbf{ v}}{dt}=-\frac{1}{\rho}\mathbf {\nabla} P = - \left[\frac{P\phi}{\rho^2} \mathbf{\nabla}\left(\frac{\rho}{\phi}\right)+\frac{1}{\phi}\mathbf{\nabla}\left(\frac{P\phi}{\rho}\right)\right]\,.
    \label{generalmomentum}
\end{equation}

The axisymmetric analog of the momentum equation is obtained from, 

\begin{equation}
\rho=\frac{\eta}{2\pi~r}\,.
\label{density}
\end{equation}

Expressing the nabla operator in cylindrical coordinates $\{r,z\}$~and differentiating the expression (\ref{density}), putting the result into Eq.~(\ref{generalmomentum}) and approaching the derivatives with summations in the usual SPH way leads to, 

\begin{equation}
\begin{split}
    \left(\frac{dv_r}{dt}\right)_a & =2\pi\frac{P_a}{\eta_a}-\\ 
    &2 \pi\sum_b  \frac{m_b}{\eta_b}\left[\frac{P_a\phi_a \vert r_a\vert}{\eta_a^2}\frac{\eta_b}{\phi_b}\frac{\partial W_{ab}}{\partial r}+\frac{\phi_b}{\phi_a}\frac{P_b\vert r_b\vert}{\eta_b}\frac{\partial W_{ab}}{\partial r}\right]\,,
    \end{split}
    \label{momentum_r}
\end{equation}

\noindent and, 
\begin{equation}
    \left(\frac{dv_z}{dt}\right)_a=
    2 \pi\sum_b  \frac{m_b}{\eta_b}\left[\frac{P_a\phi_a \vert r_a\vert}{\eta_a^2}\frac{\eta_b}{\phi_b}\frac{\partial W_{ab}}{\partial z}+\frac{\phi_b}{\phi_a}\frac{P_b\vert r_b\vert}{\eta_b}\frac{\partial W_{ab}}{\partial z}\right]\,.
    \label{momentum_z}
\end{equation}

Choosing $\phi=1$~above leads to the standard axisymmetric SPH momentum equations, which are the same as those obtained with the minimum action principle \citep{brookshaw1985}. Picking $\phi=\eta$~leads to the geometric  density averaged schemes, which are better suited to suppress the tensile instability \citep{rea10}. As shown in Sect.(\ref{sec:axisformulation}), both families of equations can be reduced to a single expression steered by a binary parameter $\sigma[0,1]$. The ensuing momentum equations are those given by Eqs.~(\ref{accel_r}) and (\ref{accel_z}). Note that the radial component of the acceleration, Eq.~(\ref{momentum_r}), has a term which does not depend of the gradient of the kernel. Such term, called the hoop-stress, is an outstanding feature of the axisymmetric geometry.    

A suitable expression for the energy equation is, 

\begin{align}
\left(\frac{du}{dt}\right)_a=-2\pi\frac{P_a}{\eta_a} v_{r_a}+
&2\pi\frac{P_a\phi_a\vert r_a\vert}{\eta_a^2}\sum_{b=1}^N \frac{m_b}{\phi_b} {\bf v}_{ab}\cdot {\bf\mathcal D }W_{ab}(h_a)\,,
\label{energyA1}
\end{align}

\noindent where ${\bf {\mathcal D}}=\frac{\partial}{\partial r} \hat {\bf r}+ \frac{\partial}{\partial z} \hat {\bf z} $~is the 2D-axisymmetric form of the ${\bf \nabla}$~operator. Making use of the parameter $\sigma[0,1]$, Equation (\ref{energyA1}) is written as Eq.(\ref{energy1}) in Sect.~(\ref{sec:axisformulation}) 
    
 \section{Derivation of the axisymmetric SPHMHD equations}
\label{sec:appendix B}

A rather common, and perhaps the most natural procedure to formulate the SPHMHD equations, is to make use of the variational  principle $\delta S =\int \delta L dt=0$~\citep{pri2004,pri12},  where the physical action, $S$, is minimized. In the following we {\sl closely follow} the demonstration in \cite{pri12}, but adapting it to the peculiarities of axial symmetry, and we refer the reader to that paper for the details. The variation of the Lagrangian of the system, $\delta L$, including the magnetic energy is,

\begin{equation}
    \begin{split}
    \delta L = & m_a v_a \cdot \delta v_a-\\&\sum_b m_b \left[\left(\frac{\partial u_b}{\partial\rho_b}\delta\rho_b\right)+\frac{1}{2\mu_0}\left(\frac{B_b}{\rho_b}\right)^2\delta\rho_b+ \frac{1}{\mu_0}{\mathbf B}_b\cdot\delta\left(\frac{{\mathbf B}_b}{\rho_b}\right)\right]\,,
    \end{split}
    \label{deltaLagrangian}
\end{equation}
  
Using Eq.~(\ref{density1}), the density variation in axial geometry is, 

\begin{equation}
    \begin{split}
    \delta\rho_b=&-\frac{\eta_b}{2\pi r_b^2}\delta {\bf s}_b+\frac{1}{2\pi r_b}\delta\eta_b
    =\\&-\frac{\eta_b}{2\pi r_b^2}\delta {\bf s}_b+\frac{1}{2\pi r_b}\sum_c m_c\left(\delta{\mathbf s}_b-\delta{\mathbf s}_c\right)\cdot {\mathbf{\mathcal D}}_b W_{bc}(h_b)
    \end{split}
    \label{deltarho}
\end{equation}

Direct substitution of  $\delta \rho_b$~above, besides $\frac{\partial u_b}{\partial\rho_b}=\frac{P_b}{\rho_b^2}$~and $\rho_b=\frac{\eta_b}{2\pi r_b}$~in Eq.~(\ref{deltaLagrangian}) leads to identical contributions to the acceleration as in \cite{pri12}, except those arising from the hoop-stress force:

\begin{equation}
\left(a^r_{hoop}\right)_b=2\pi\frac{P_b+\frac{B_b^2}{2\mu_0}}{\eta_b}
\label{hoop1}
\end{equation}

We now evaluate the contribution of the last term on the RHS in Eq.~(\ref{deltaLagrangian}) separately. The magnitude $\delta (\frac{{\mathbf B}_b}{\rho_b})$ is obtained from the magnetic induction equation, 

\begin{equation}
\frac{d}{dt}\left(\frac{{\bf B}}{\rho}\right) = \left(\frac{\mathbf B}{\rho}\cdot \nabla\right){\mathbf v}\,,
\label{induction2}
\end{equation}
    
\noindent which, once expressed in cylindrical coordinates, and with $\partial/\partial\varphi=0$, becomes,

\begin{equation}
\begin{split}
    \frac{d}{dt}\left(\frac{{\bf B}}{\rho}\right)=&\left(\frac{B_r}{\rho}\frac{\partial v_r}{\partial r}+\frac{B_z}{\rho}\frac{\partial v_r}{\partial z}-\frac{B_{\varphi} v_{\varphi}}{\rho r}\right)\hat r + \\
    & \left(\frac{B_r}{\rho}\frac{\partial v_z}{\partial r}+ \frac{B_z}{\rho}\frac{\partial v_z}{\partial z}\right)\hat z+\\
    & \left(\frac{B_r}{\rho}\frac{\partial v_\varphi}{\partial r}+\frac{B_z}{\rho}\frac{\partial v_\varphi}{\partial z}+\frac{B_\varphi v_r}{ \rho r}\right)\hat\varphi\,.
    \end{split}
    \label{induction3}
\end{equation}
    
The expression above has contributions from the velocity of the particle and from the derivatives of the velocity (noted next with superscript $Dvel$). With SPH summations, the latter are,

\begin{equation}
    \frac{d}{dt}\left(\frac{{\bf B}}{\rho}\right)^{Dvel}_b=-\sum_c m_c ({\mathbf v}_c-{\mathbf v}_b)\left \{\frac{{\mathbf B}_b}{\rho_b^2}\cdot {\mathbf{\mathcal D}}_b W_{bc}(h_b)\right\}\,,
    \label{induction4}
\end{equation}
    
\noindent and,     
 
 \begin{equation}
    \delta\left(\frac{{\bf B}}{\rho}\right)^{Dvel}_b=-\sum_c m_c ({\mathbf {\delta R}}_c-{\mathbf {\delta  R}}_b)\left \{\frac{{\mathbf B}_b}{\rho_b^2}\cdot {\mathbf{\mathcal D}}_b W_{bc}(h_b)\right\}\,,
    \label{induction5}
\end{equation}

\noindent where ${\mathbf {\delta R}\equiv (\delta r,\delta z, \delta\varphi)}$~stands for a three-dimensional virtual displacement (but note that $\mathcal {\mathbf D} \equiv (\frac{\partial}{\partial r},\frac{\partial}{\partial z}, 0)$~is the nabla-operator restricted to the axisymmetric plane). The scalar product, 

\begin{equation}
  \frac{\mathbf {B}}{\mu_o}\cdot  \delta\left(\frac{{\bf B}}{\rho}\right)^{Dvel}\,,
  \label{induction6}
\end{equation}

 \noindent is formally the same as that in \cite{pri12}, with one coordinate changed to $\varphi$,  thus leading to a similar contribution to the acceleration via the variational principle. It is worth noting that if the azimuthal velocity $v_\varphi$~is not zero, or has non-zero derivatives, it induces an acceleration $a_\varphi$~orthogonal to the axisymmetric plane. 
 
 Finally, the terms $-\frac{B_\varphi v_\varphi}{\rho r}$~and $\frac{B_\varphi v_r}{\rho r}$~in Eq.~(\ref{induction3}) lead to additional hoop-stress contributions to the accelerations $a^\varphi$~and $a^r$~respectively, 
 
 \begin{equation}
     a^r_{hoop}= -\frac{B_\varphi^2}{\mu_0\rho r}= -2\pi \frac{B_{\varphi}^2}{\mu_0 \eta}\,,
     \label{hoop2}
 \end{equation}
 
 \noindent which comes from a virtual displacement along the $r-$direction and has to be added to Eq.~(\ref{hoop1}) to compute the total hoop-stress force. Similarly, a virtual displacement in the $\varphi$-direction leads to,
 
 \begin{equation}
     a^\varphi_{hoop}=\frac{B^r B^\varphi}{\mu_0\rho r}= 2\pi\frac{B^r B^\varphi}{\mu_0\eta}\,,
 \end{equation}
 
 \noindent which in some particular scenarios will make its way to  contribute to the tangential acceleration $a^\varphi$~of the particle.
 
 To summarize, the different components of the momentum equation, written in ISPH notation, read:
 
  \begin{equation}
\begin{split}
\left(\frac{dv^r}{dt}\right)_a=~ & 2\pi\frac{\left(P_a+\frac{B_a^2}{2\mu_0}-\frac{\left(B_a^\varphi\right)^2}{\mu_0}\right)}{\eta_a}-\\
&2\pi \sum_{b=1}^{n_b}  m_b \left(\frac{S_a^{ri} \vert r_a\vert}{\eta^2_a}{\mathcal A}_{ab}^{i}(h_a)+
 \frac{S_b^{ri} \vert r_b\vert}{\eta_b^2} {\mathcal A}_{ab}^{i}(h_b)\right)\,, 
\end{split}
\label{App_mhdaccel_r}
\end{equation}

\begin{equation}
\left(\frac{d v^z}{dt}\right)_a=  2\pi\sum_{b=1}^{n_b} m_b\left(\frac{S_a^{zi} \vert r_a\vert}{\eta^2_a}{\mathcal A}_{ab}^{i}(h_a)+\frac{S_b^{zi} \vert r_b\vert}{\eta^2_b}{\mathcal A}_{ab}^{i}(h_b)\right)\,.
\label{App_mhdaccel_z}
\end{equation}

The momentum equation in the $\hat\varphi$~direction arises from the last term on the RHS in Eq.~(\ref{deltaLagrangian}),  

\begin{equation}
\begin{split}
\left(\frac{dv^{\varphi}}{dt}\right)_a=~ &  2\pi\left(\frac{B_a^r B_a^{\varphi}}{\mu_0\eta_a}\right)+\\
&2\pi \sum_{b=1}^{n_b}  m_b \left(\frac{S_a^{\varphi i} \vert r_a\vert}{\eta^2_a}{\mathcal A}_{ab}^{i}(h_a)+
 \frac{S_b^{\varphi i} \vert r_b\vert}{\eta_b^2} {\mathcal A}_{ab}^{i}(h_b)\right)\,,
\end{split}
\label{App_mhdaccel_phi}
\end{equation}

\noindent with $i=r,z$~in all equations, and repeated indexes are summed up. After minor modifications, to account for axis corrections and to reduce the magnetic tensile instability, equations (\ref{App_mhdaccel_r}) and (\ref{App_mhdaccel_z}) turn into Eqs.~(\ref{mhdaccel_r}) and (\ref{mhdaccel_z}) described in Sect.~\ref{sec:isphmhd}. These take over the evolution in the tests described in Sects.~\ref{subsec:mloop} to \ref{subsec:KH}. The equation (\ref{App_mhdaccel_phi}) is useful to simulate magnetized systems with non-zero initial angular momentum, such as the rotating cloud test described in Sect.~\ref{subsec:collapse}.

\section{Axisymmetric form of the magnetic dissipation}
\label{sec:appendix C}

Some amount of magnetic dissipation is necessary to handle shock waves. We make use of the expression by \cite{price18}, 

\begin{equation}
    \left(\frac{d\mathbf B}{dt}\right)^{diss}=\xi_B~\nabla^2\mathbf B\,
    \label{ApBdis_1}
\end{equation}

\noindent where $\xi_B$ is a resistivity parameter. Equation (\ref{ApBdis_1}) is adapted to the axisymmetric geometry by writing the vector Laplacian on the RHS in Eq.~(\ref{ApBdis_1}) in cylindrical coordinates \footnote{https://mathworld.wolfram.com/VectorLaplacian.html}, with the constraints  $\partial/\partial\varphi=0, \partial^2/\partial\varphi^2=0$, 

\begin{equation}
    \begin{split}
    \nabla^2 {\bf B}=&\left[\left(\frac{\partial^2B^r}{\partial r^2}+\frac{\partial^2 B^r}{\partial z^2}\right)+\frac{1}{r}\frac{\partial B^r}{\partial r}-\frac{B^r}{r^2}\right]\hat {\bf r}~+
    \\& \left[\left(\frac{\partial^ 2B^z}{\partial r^2}+\frac{\partial^2 B^z}{\partial z^2}\right)+\frac{1}{r}\frac{\partial B^z}{\partial r}\right]\hat {\bf z}~+\\
    &\left[\left(\frac{\partial^2 B^\varphi}{\partial r^2}+ \frac{\partial^2 B^\varphi}{\partial z^2}\right)+\frac{1}{r}\frac{\partial B^\varphi}{\partial r}-\frac{B^\varphi}{r^2}\right]\hat {\bf \varphi}\,.
    \end{split}
    \label{ApBdis_2}
\end{equation}

The second derivatives in parenthesis in the RHS of Eq.(\ref{ApBdis_2}) are the Cartesian 2D-Laplacian of each component of the magnetic field, ${\mathcal D}^2 B^i $, which are computed in the standard SPH way \cite{},   

\begin{equation}
    \left(\xi_B~{\mathcal D}^2 B^i\right)_a =\sum_{b=1}^{n_b} V_b~ \frac{\xi_{B,a}+\xi_{B,b}}{\vert s_{ab}\vert} B^i_{ab}\left(\hat s^j_{ab}\tilde {\mathcal A}^j_{ab}\right)\,.
    \label{ApBdiss_3}
\end{equation}

The first derivatives at the RHS of Eq.(\ref{ApBdiss_3}) are estimated with, 

\begin{equation}
   \left(\xi_B~\frac{\partial B^i}{\partial r}\right)_a=\xi_{B,a}\sum_{b=1}^{n_b} V_b \left(B_a^i-B_b^i\right){\mathcal A}^j_{ab} (h_a)\,.
\end{equation}

The expression giving the magnetic dissipation in axial geometry is therefore written, 

\begin{equation}
\left(\frac{dB^i}{dt}\right)_a^{diss} =\left(\xi_B~{\mathcal D}^2 B^i\right)_a+\left(\xi_B\frac{1}{r}~\frac{\partial B^i}{\partial r}\right)_a - (1-\delta^{i2}) \left(\xi_B\frac{B^i}{r^2}\right)_a\,,
\end{equation}

\noindent where $i=\{1,2,3\}$~correspond to components  $\{r,z,\varphi\}$~and $\delta^{i2}$~is the Kronecker-delta.  

\bsp	
\label{lastpage}
\end{document}